\documentclass[]{article}

\usepackage{graphics} % for pdf, bitmapped graphics files
\usepackage{epsfig} % for postscript graphics files
\usepackage{mathptmx} % assumes new font selection scheme installed
\usepackage{times} % assumes new font selection scheme installed
\usepackage{amsmath} % assumes amsmath package installed
\usepackage{amssymb}  % assumes amsmath package installed
\usepackage{graphicx}
\usepackage{epsfig,epstopdf,color}

\begin{document}

\title{Formally expressing the semantics of observer-based fault detection software}

\author{Alireza~Esna~Ashari\thanks{A. Esna Ashari is with Georgia Institute of Technology, USA.
{\tt\small esna\_ashari@gatech.edu}}~~and~Eric~Feron%
\thanks{E. Feron is with is with Georgia Institute of Technology, USA.
 {\tt\small feron@gatech.edu}}}

\maketitle

\newtheorem{theorem}{Theorem}[section]
\newtheorem{lemma}[theorem]{Lemma}
\newtheorem{proposition}[theorem]{Proposition}
\newtheorem{corollary}[theorem]{Corollary}
\newtheorem{rem}[theorem]{Remark}

\begin{abstract}
The aim is to create reliable and verifiable fault detection software to detect abrupt changes in safety-critical dynamic systems. Fault detection methods are implemented as software on digital computers that monitor and control the system.
We implement three observer-based fault detection methods on a 3 degrees of freedom (3DOF) laboratory helicopter, in the form of software.
We examine the performance of those methods to detect different faults during flight in a closed-loop setup.
All selected methods show acceptable detection performance. However, it is not possible to repeat the test for every possible conditions, inputs and fault scenarios. In this paper, we translate fault detection properties and mathematical proofs into a formal language, previously used in software validation and verification. We include the translated properties in software in the form of non-executable annotations that can be read by machine. Consequently, some high level functional properties of the code can be verified by automatic software verification tools. This certifies fault detection software for a set of bounded data and increases the reliability in practice.
\end{abstract}


\section{Introduction}

Fault detection is an essential part of any complicated and safety-critical system. For years, several methods have been introduced for detecting possible issues in dynamic systems to guarantee normal functionality of the system. In practice, the designer selects one out of several fault detection methods, based on the specifications of the system and the nature of possible faults. Some methods are more suitable for off-line fault detection test. One example is subspace-based detection method, introduced in \cite{basse-aut00,esna2012input}. The method is used for health monitoring of mechanical structures, such as bridges. Other methods aim at detecting faults online. All observer-based methods belong to this category. The observer or estimator filter provides analytical redundancy for health monitoring of the system, analyzing input-output data and comparing the result with the nominal behavior of the system. Most studies on model-based fault detection develop such approaches. A survey on such methods is provided in \cite{hwang2010survey} (see also \cite{gertler1988survey,frank1990fault,frank1997survey,isermann1997supervision}). Among those approaches, some are stochastic and the others are deterministic. All those methods are passive in the sense that the detector only monitors input and output data and decides on health of the system.
Recently, active methods are introduced that aim at guaranteeing or improving fault detection by designing and injecting auxiliary signals into the system (see \cite{campbell2004auxiliary,esna2011auxiliary,niemann1997robust,niemann2005architecture}). Among all the aforementioned methods, some can be used only in open-loop setup, while the others can be applied to closed-loop systems. Fault isolation and estimation is another property that only some of the methods provide (see \cite{ding2008model,patton1999robust,isermann2005fault}). In conclusion, the designer has a broad range of selection and can choose the most suitable method based on all the requirements.

After selecting a proper fault detection method, the designer needs to calculate the design parameters and simulate the method, based on system model, and using software such as Matlab, Simulink, Scicos, etc. A successful simulation result is promising to move on to the next step. Then the method is implemented on the real system. In recent years, digital computers and micro controllers have been developed rapidly. Now, it is economically reasonable to implement control, estimation and fault detection methods in the form of software on digital computers. Therefore, software that already passed simulation tests can be installed on computers or micro controllers connected to the real system. If the real system is complicated and sensitive, a prototype, small model or a physical simulator can be used to test the method before implementing to the real system.

Through the process of design and implementation, several scientists and engineers may collaborate together. The software implementation of control, estimation and fault detection methods can be finalized or be modified by computer scientists and the code created initially by a control engineer might be merged to the other pieces of codes. In the end, the software might be different from what it is supposed to be in the original design, can have bugs and errors and may not satisfy the expectations of the original designer. Some parameters might be changed by trial and error and, finally, there is no mathematical evidence supporting the flawless operation of the implemented method. Even if the code does not undergo any changes, computers work with binary numbers and floating point that computer use are different from real numbers in math. Therefore, there always exist calculation errors and the implemented software is deviated from the theory that supports it. In conclusion, the results are not exactly what we expect from control and signal processing theories. This issue is more crucial for fault detection in safety-critical systems, such as airplanes and helicopters. In the light of evolving software certification requirements, it becomes important to formally specify the correctness of these algorithms. We look for the solution in software verification.

Software verification is already implemented to check the stability of the control systems \cite{feron2010control} (see also \cite{feron2007certifying,roozbehani2011optimization}). The aim of control software verification tools is to verify the correct implementation of control methods in the form of codes. For that purpose, software annotations are included in control software so that the software can be readable for computer scientists and be analysed by verification tools. Control properties are expressed by invariant sets to which software variables belong. It makes a bridge for the other scientists who are not familiar with control or signal processing techniques to verify the software. The final aim is to verify the software using verification tools that can read the annotations, evaluate formal methods and
certified the proofs at the level of the software (see \cite{Zapana2012}).

In this research we select three observer-based fault detection methods, capable of detecting abrupt changes in closed-loop systems. We implement the methods in the form of software to monitor a laboratory 3DOF helicopter and show that all the methods can detect some pre-specified fault scenarios. However, it does not guarantee the functionality of the software in any circumstances. Therefore, the question is how to develop formal methods of verification for the software and create a detection software that provides us a higher level of reliability. Properties that have mathematical proofs are translated in formal language, understandable for computer scientists, as well as computers that verify software automatically, and are added to the software as annotations. Hence, those specifications are preserved in software level and can contribute to the validation of a complete system during implementation process.

%%%%%%%%%%%%%%%%%%%%%%%%%%%%%%%%%%%%%%%%%%%%%%%%%%%%%%%%%%%%%%%%%%%%%%%%%%%%%%%%
\section{System modeling and problem formulation}

Model-based fault detection methods need the system to be represented by mathematical equations.
Continuous-time dynamic systems can be modeled by differential equations. Original models are usually nonlinear, but a linear model of the following form
\begin{eqnarray}
\dot{x}(t)&=&A x(t)+B u(t)+ E_d f_d(t)+ E_f f(t), \label{1}\\
y(t)&=&C x(t), \label{2}
\end{eqnarray}
can be obtained around the operating point of the system, where $x(t)\in \Re^n$ and $u(k) \in \Re^m$ are the state vector and the known input vector at time $t$, respectively. $f(t) \in \Re^{n_f}$ is an additive fault to the system that should be detected. $f_d(t) \in \Re^{n_d}$  represents all unknown signals to the system that we do not aim at detecting. This signal acts as a secondary fault to the system. No prior knowledge on $f(t)$ and $f_d(t)$ is available. Although the methods in this paper are capable of detecting sensor faults, we do not discuss sensor fault detection in that context to avoid complexity. Instead we focus on the correct software implementation.
$A \in \Re^{n\times n}$ and $B \in \Re^{n\times m}$ are state transition and input matrices, respectively.
Here, $y(t) \in \Re^{p}$ is the output vector and  $C \in \Re^{p \times n}$ is the corresponding output matrix. We assume $(A,C)$ is observable. $E_f$ determines how the system fault affects the system and $E_d$ defines the structure of faults that we do not want to detect.
In this paper we consider three problems, namely fault detection, fault detection and isolation and fault estimation.
The value of $f(t)$ and $f_d(t)$ are zero for nominal (fault-free) system. The aim of the fault detection is to raise an alarm whenever these value differs significantly from zero (faulty system). Note that most unwanted changes in the system that lead to a system matrix change can be represented by signal $f(t)$. Fault detection and isolation considers the case in which the observer detects only particular faults and bypasses the others.
We are only interested in detecting $f(t)$.
The detector must be as sensitive as possible to $f(t)$ and, at the same time, insensitive to $f_d(t)$.
Finally fault estimation provides a method to reconstruct $f(t)$ and $f_d(t)$.

\section{Observer-based fault detection: informal methods} \label{theory}

We consider three different observer-based fault detection methods in this paper. In this section, we briefly
summarize the theory of those methods. Those methods have been used for designing fault detection observers for years. However, the original methods are not  expressed in the formal language of computer science and software verification. We call them informal methods to comply with computer science terminology.
In Section \ref{veri} we translate the properties of these methods into formal language.

\subsection{Output observer design for fault detection} \label{output}

The simplest observer that can be used for fault detection is the output observer. It monitors the outputs of the real system and compares them with those simulated with a nominal model of the system.
The material in this section has been explained in detail in most fault detection books, see for example \cite{ding2008model,isermann2005fault}. In this approach, no specific strategy is provided for fault isolation. Therefore the observer is sensitive  to both $f(t)$ and $f_d(t)$, and detects both.
%  Note that in practice, one may select design parameters by trial and error so that isolability is achieved. However, the method does not guarantee the isolability.
The core of the output observer detector is a full-order state observer
\begin{eqnarray}
\dot{\hat{x}}(t)&=&A \hat{x}(t)+B u(t)+ L (y(t)-C \hat{x}(t)), \label{3}\\
\hat{y}(t)&=&C \hat{x}(t). \label{3a}
\end{eqnarray}
A simple output observer \eqref{3} produces a residual signal, $r(t)$, to compare the estimated output with the measured one
\begin{equation}
r(t)=y(t)-\hat{y}(t). \label{4}
\end{equation}
Introducing the estimation error, $e(t)=x(t)-\hat{x}(t)$, we calculate the dynamics of the error
\begin{eqnarray}
\dot{e}(t)&=&(A-LC)e(t)+E_d f_d(t)+E_f f(t), \label{5}\\
r(t)&=&Ce(t). \label{5a}
\end{eqnarray}
From \eqref{5}--\eqref{5a}, $r(t)$ goes to zero for the nominal system (where $f(t)$ and $f_d(t)$ are zero) when
the observer matrix $L$ is so chosen that $A-LC$ is stable. In nominal case, $\hat{x}(t)$ also
provides a unbiased estimate for $x(t)$, i.e.
\begin{equation}
\lim_{t \rightarrow \infty} (x(t)-\hat{x}(t))=0. \label{6}
\end{equation}
For this simple observer, the only design parameter is $L$, which should be selected so that
the error dynamics is stable. We can use any pole placement method to select $L$ for that purpose.
When the observer is implemented as software, it receives input-output data, the residual signal $r(t)$ is calculated online and the norm of this signal is usually compared to a threshold. If the norm of $r(t)$ reaches the threshold a fault alarm is raised. We discuss about it more in Section \ref{output-steady}. The threshold is selected so that small norm faults does not raise an alarm.

\subsection{Fault detection and isolation using Unknown Input Observers (UIO)} \label{UIO}

The method introduced in \cite{patton1999robust} is summarized in this section, with slight modifications.
An unknown input observer for the system \eqref{1}--\eqref{2}  is an observer that its state estimation error approaches zero asymptotically, regardless of the presence of the unknown input $f_d(t)$ in the system. Several methods have been proposed in the literature to design UIO. The order of the observer might be different from that of the system \cite{patton1999robust}. Here we assume that the observer and the system are of the same order.

The observer is modeled as
\begin{eqnarray}
\dot{z}(t)&=&F z(t)+T B u(t)+ K y(t), \label{12}\\
\hat{x}(t)&=&z(t) + H y(t), \label{13}\\
r(t)&=&y(t)-C\hat{x}(t). \label{13b}
\end{eqnarray}
It has been proved that UIO does not exist if
\begin{eqnarray}
rank(CE_d)\neq rank(E_d). \label{13c}
\end{eqnarray}
Notice that if the number of fault input channels is bigger than
the number of measured outputs, we cannot design a UIO. Suppose that \eqref{13c} is not true.
Assume that $e(t)=\hat{x}(t)-x(t)$ is the estimation error.  Let $FH$ be represented by $K_2$ and $K=K_1+FH$. Hence we have $K=K_1+K_2$. The error dynamics can be obtained as follows
\begin{eqnarray}
\dot{e}(t)&=&(A-HCA-K_1C)e(t) \nonumber \\
&+& [F-(A-HCA-K_1C)]z(t) \nonumber \\
&+& [K_2-(A-HCA-K_1C)H]y(t) \nonumber \\
&+& [T-(I-HC)]Bu(t)+(HC-I)E_d f_d(t) \nonumber \\
&+& (HC-I)E_f f(t). \label{14}
\end{eqnarray}
In order to have stable error dynamics, we enforce the following conditions
\begin{eqnarray}
& & (HC-I)E_d=0,\nonumber \\
& & T=I-HC, \nonumber\\
& & F=A-HCA-K_1C, \label{15}
\end{eqnarray}
and select $F$ to be Hurwitz. Such a selection leads to $\dot{e}(t)=F e(t)$ for non-faulty system
($f(t)$ is zero) and the observation error asymptotically converges to zero. From \eqref{15}, all we need to stabilize $F$ is to select $K_1$ using pole placement methods. When
$(C , A-HCA)$ is observable, $K_1$ can be calculated using conventional eigenvalue assignment approaches. No more algebraic manipulations are required.

If $(C , A-HCA)$ is not observable but detectable, a particular approach can be used to design $K_1$. In what follows, we explain the approach. Note that $(C , A-HCA)$ is detectable whenever it is either observable or non-observable modes are stable.

We can find a similarity transformation matrix $P$ so that (see \cite{rosenbrock1970state,patton1999robust})
\begin{eqnarray}
P(A-HCA)P^{-1} &=& \begin{pmatrix} A_{11} & A_{12} \\ 0 & A_{22}\end{pmatrix}, \label{16}\\
CP^{-1}&=& \begin{pmatrix} 0 & C_{2} \end{pmatrix} .\label{17}
\end{eqnarray}
In that case $A_{11}$ is stable if $(C , A-HCA)$ is detectable. We select matrix $K_{p}^{2}$ so that the eigenvalues of $A_{22}-K_{p}^{2}C_2$ are assigned at desired stable locations. Then $K_1$ can be obtained as
\begin{eqnarray}
K_{1}= P^{-1} \begin{pmatrix} K_p^1 \\ K_p^2 \end{pmatrix}, \label{18}
\end{eqnarray}
where $K_p^1$ is any arbitrary matrix of proper dimensions. In the end, observer matrices are computed as follows
\begin{eqnarray}
H&=&E_d[(CE_d)^TCE_d]^{-1}(CE_d)^T,\;\; T=I-HC, \nonumber \\
F&=& A-HCA -K_1C, \;\;  K=K_1+FH.\label{19}
\end{eqnarray}
The designed UIO is robust to unknown input $f_d(t)$. Hence, we have
\begin{eqnarray}
\dot{e}(t)&=&F e(t)-TE_f f(t) ,\label{20}\\
r(t)&=&Ce(t).\label{21}
\end{eqnarray}
If $TE_f\neq0$, the additive fault $f(t)$ can be detected.

\subsection{Fault estimation using sliding-mode fault detector} \label{sliding}

Consider again the system \eqref{1}--\eqref{2}. It can be rewritten as
\begin{eqnarray}
\dot{x}(t)&=&A x(t)+B u(t)+ \bar{E}_f \bar{f}(t), \label{22}\\
y(t)&=&C x(t), \label{23}
\end{eqnarray}
where
\begin{eqnarray}
\bar{E}_f = \begin{pmatrix} E_f & E_d \end{pmatrix}, \label{59}\\
\bar{f}(t)= \begin{pmatrix} f(t) \\ f_d(t) \end{pmatrix}. \label{60}
\end{eqnarray}
The sliding-mode fault estimator is supposed to estimate $\bar{f}(t)$. Notice that fault estimation is more complicated than fault isolation. It calculates an estimation of any unknown input signal.
We assume that $B$, $\bar{E}_f$ and $C$ are full rank and the number of system faults is smaller than the number of outputs. Also we suppose that any invariant zero of $(A,\bar{E}_f,C)$ is stable. These assumptions guarantee the existence of a sliding-mode detector. An sliding-mode detector design approach is introduced in \cite{edwards1998sliding}.
The following observer is proposed to estimate the states of the system \eqref{22}--\eqref{23}
\begin{eqnarray}
\dot{\hat{x}}(t)&=&A \hat{x}(t)+B u(t)+ G_l e_y(t)+G_n \nu,  \label{24} \\
\hat{y}(t)&=&C\hat{x}(t), \label{25}\\
e_y(t)&=&\hat{y}(t)-y(t), \label{26}
\end{eqnarray}
where,
\begin{eqnarray}
G_n&=&T_o^{-1} \begin{pmatrix} 0 \\I \end{pmatrix}, \label{27}\\
G_l&=&T_o^{-1} \begin{pmatrix} A_{12} \\A_{22}-A_{22}^s \end{pmatrix},\label{28}
\end{eqnarray}
and
\begin{eqnarray}
\nu= \left\{
  \begin{array}{l l}
    -\rho(t,y,u) \|D_2\| \frac{P_2 e_y}{\|P_2 e_y \|+\sigma} & \quad \textrm{if $e_y\neq0$} \\
    0 & \quad \textrm{otherwise.}
  \end{array} \right. \label{29}
\end{eqnarray}
Here, $D_2$, $A_{12}$, $A_{22}$, $A_{22}^s$, $\rho(t,y,u)$,  $P_2$ and $T_o$ must be selected so that $e_y$ goes to zero in finite time. $\sigma$ is a small positive scalar added to the denominator of the discontinuous part of the $\nu$ to reduce the chattering effects \cite{edwards1998sliding}.
The corresponding sliding surface is $\{e\in \Re^n |Ce=0\}$. Also we suppose $\rho(t,y,u)$ is selected so that \cite{edwards2000sliding}
\begin{eqnarray}
\|\bar{f}(t)\| < \rho(t,y,u).\label{51}
\end{eqnarray}
In practice, $\rho$ can be selected as a positive constant scalar \cite{edwards1998sliding}.
Here, we briefly explain the design procedure. It can be shown that there exists a similarity transformation matrix, $T_o$, that can transform system \eqref{22}--\eqref{23} into
\begin{eqnarray}
\dot{x}_1(t)&=&A_{11} x_1(t)+A_{12} x_2(t)+ B_1 u(t), \label{30}\\
\dot{x}_2(t)&=&A_{21} x_1(t)+A_{22} x_2(t)+B_2 u(t) + D_2 \bar{f}(t), \label{31}\\
y(t)&=&x_2(t), \label{32}
\end{eqnarray}
where $A_{11}$ has negative eigenvalues.
For \eqref{30}--\eqref{31} we use the observer bellow
\begin{eqnarray}
\dot{\hat{x}}_1(t)&=&A_{11} \hat{x}_1(t)+A_{12} \hat{x}_2(t)+ B_1 u(t)-A_{12} e_y(t), \label{33}\\
\dot{\hat{x}}_2(t)&=&A_{21} \hat{x}_1(t)+A_{22} \hat{x}_2(t)+B_2 u(t) \nonumber\\
&-& (A_{22}-A_{22}^s)+\nu, \label{34}\\
\hat{y}(t)&=&\hat{x}_2(t), \label{35}
\end{eqnarray}
in which $A_{22}^s$ is a design matrix that must be Hurwitz. $P_2$ in \eqref{29} is a Lyapunov matrix for $A_{22}^s$ that guarantees quadratic stability of the error dynamics of the observer
\begin{eqnarray}
\dot{e}_1(t)&=&A_{11} e_1(t), \label{36}\\
\dot{e}_y(t)&=&A_{21} e_1(t)+A_{22}^s e_y(t)+\nu-D_2 \bar{f}(t), \label{37}\\
e_1(t)&=&\hat{x}_1(t)-x_1(t). \label{38}
\end{eqnarray}

Suppose that sliding motion is taken place and consequently $e_y(t)=Ce(t)=0$ and $\dot{e}_y(t)=0$. Considering \eqref{36}--\eqref{37}, and from the fact that
$A_{11}$ is stable we obtain $\lim_{t \rightarrow \infty} e_1(t) = 0$ and
\begin{eqnarray}
\nu \rightarrow D_2 \bar{f}(t)\label{39}
\end{eqnarray}
From \eqref{39}, we propose the residual
\begin{eqnarray}
r(t)= \rho \|D_2\| (D_2^T D_2)^{-1} D_2^T \frac{P_2 e_y(t)}{\| P_2 e_y(t) \|+\sigma}.\label{40}
\end{eqnarray}
Note that this residual not only does detect the fault, but also its value is an estimate of the fault.

\section{Experimental results} \label{exper}

\begin{figure}[htbp]
\begin{center}
\includegraphics
[scale=0.23]{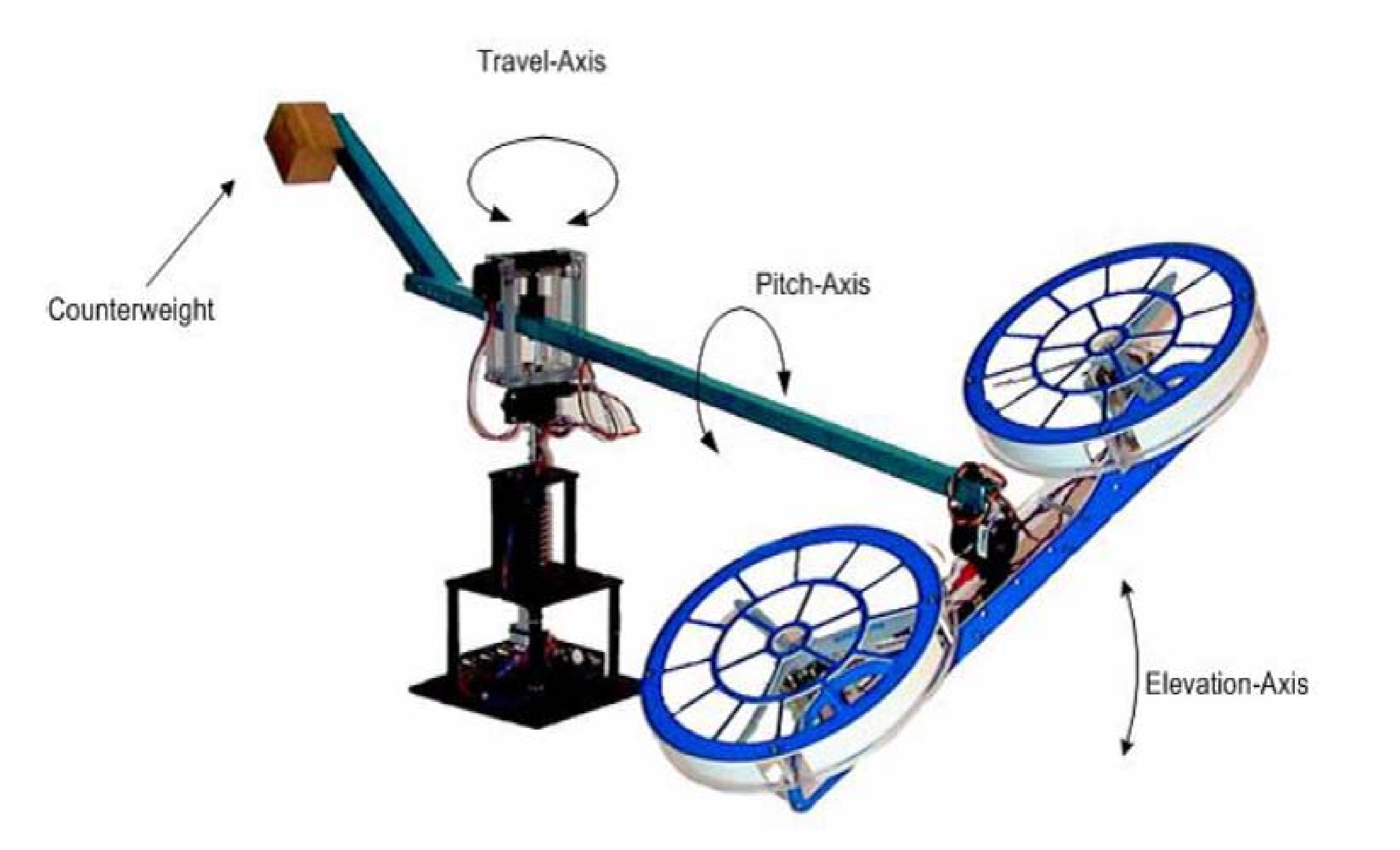}
\end{center}
\caption{Quanser 3-DOF helicopter, \cite{Quanser}}
\label{Pic_quanser}
\end{figure}

To show that the methods presented in this paper are efficient  in practice in detecting faults, we implement them in the form of software to detect some predefined faults to a laboratory 3DOF helicopter. The helicopter is shown in Figure \ref{Pic_quanser}.
A linear model of the system is provided by the manufacturer that is derived from highly nonlinear differential equations of the system. The order of the system is six and all its poles are at zero. The linear model of the system is given in Appendix. This model comes with an $LQR$ controller. In this research we use that controller. Details of controller design is given in \cite{Quanser}. Therefore we study the methods in a closed-loop setup.

\subsection{System fault in closed-loop setup} \label{scenario}

\begin{figure}[htbp]
\begin{center}
\includegraphics
[scale=0.23]{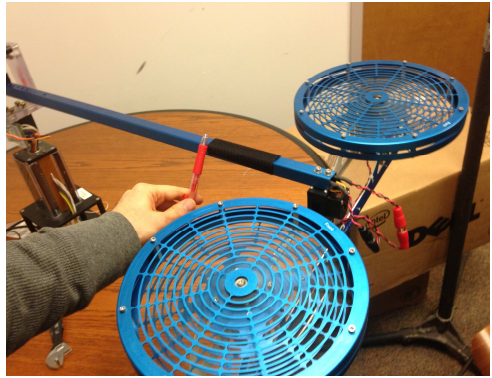}
\end{center}
\caption{First fault in the first experiment: horizonal force to the system}
\label{Pic_Horizonal}
\end{figure}

\begin{figure}[htbp]
\begin{center}
\includegraphics
[scale=0.23]{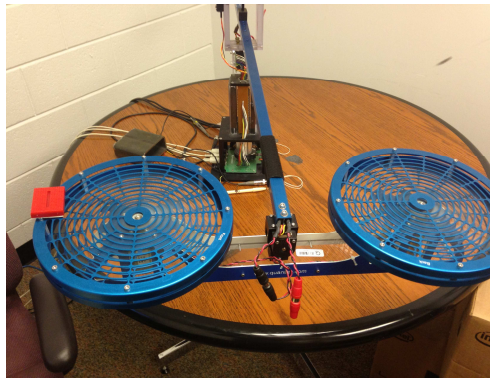}
\end{center}
\caption{Second fault in the first experiment: small mass on one of the motors}
\label{Pic_SmallMass}
\end{figure}

\begin{figure}[htbp]
\begin{center}
\includegraphics
[scale=0.23]{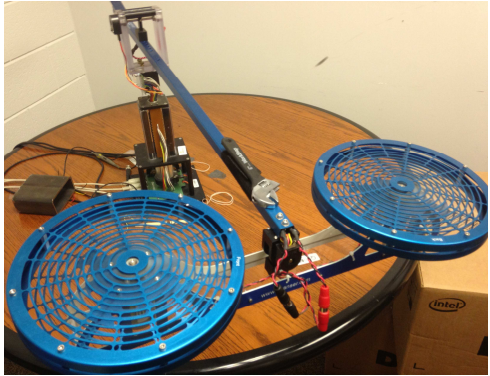}
\end{center}
\caption{Third fault in the first experiment: heavy mass on the main arm}
\label{Pic_Wrench}
\end{figure}

In general, feedback attenuate the effect of a fault in the system (see \cite{campbell2004auxiliary,esna2010active,esna2012effects}).
Therefore, it is harder to detect faults in a closed-loop setup.
The controller is already designed, and we access the input data that the controller sends to the helicopter, as well as the outputs of the system measured by sensors.
In the first experiment, three independent faults affect the system consecutively. Each fault lasts 10 seconds and then the system returns to its normal operating condition for 10 seconds. Fault 1 is a horizonal force implemented to the helicopter that deviates it from the operating point (Figure \ref{Pic_Horizonal}). It mainly changes travel angle, however it affects pitch angle inevitably. A small mass is put on one of the propellers to make Fault 2 (Figure \ref{Pic_SmallMass}). This small mass does not change the elevation angle but alters the pitch angle. Finally, Fault 3 is made by a heavy mass which is put on the main arm. This fault changes the elevation angle (Figure \ref{Pic_Wrench}).

The system operates in nominal mode in the beginning of the experiment. After 10 seconds Fault 1 takes place. This fault lasts for 10 seconds and then the system returns to the nominal mode for 10 seconds. Fault 2 happens at t=30 sec. The system returns to the nominal mode again for 10 seconds. At t=50 sec. fault 3 takes place and lasts till the end of the experiment.

\begin{figure}[htbp]
\begin{center}
\includegraphics
[scale=0.23]{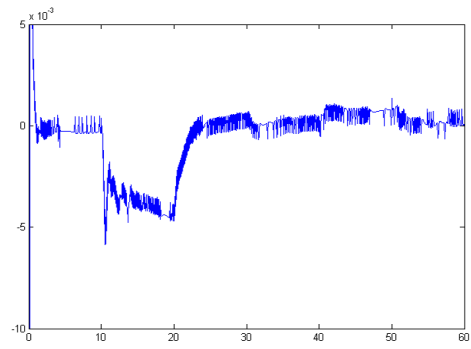}
\end{center}
\caption{Output observer method: first residual detects the first fault}
\label{Pic_r1}
\end{figure}

\begin{figure}[htbp]
\begin{center}
\includegraphics
[scale=0.23]{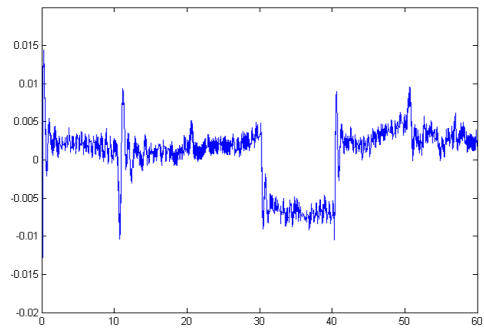}
\end{center}
\caption{Output observer method: second residual detects the second fault}
\label{Pic_r2}
\end{figure}

\begin{figure}[htbp]
\begin{center}
\includegraphics
[scale=0.23]{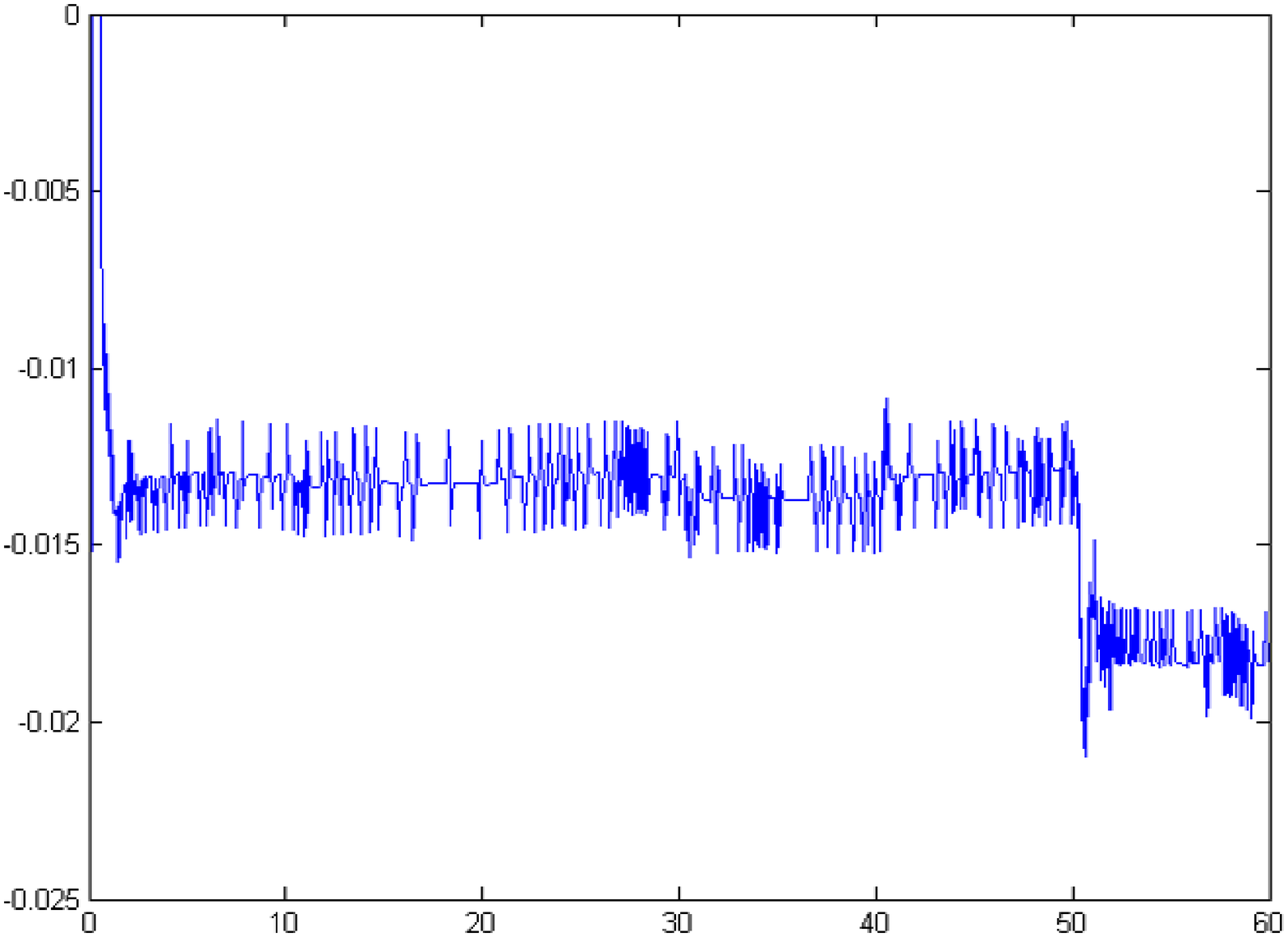}
\end{center}
\caption{Output observer method: third residual detects the third fault}
\label{Pic_r3}
\end{figure}

This experiment is repeated using the three fault detection methods. Figures \ref{Pic_r1}--\ref{Pic_r3} show the 3 residual signals produced by the output observer (first method).

\begin{figure}[htbp]
\begin{center}
\includegraphics
[scale=0.23]{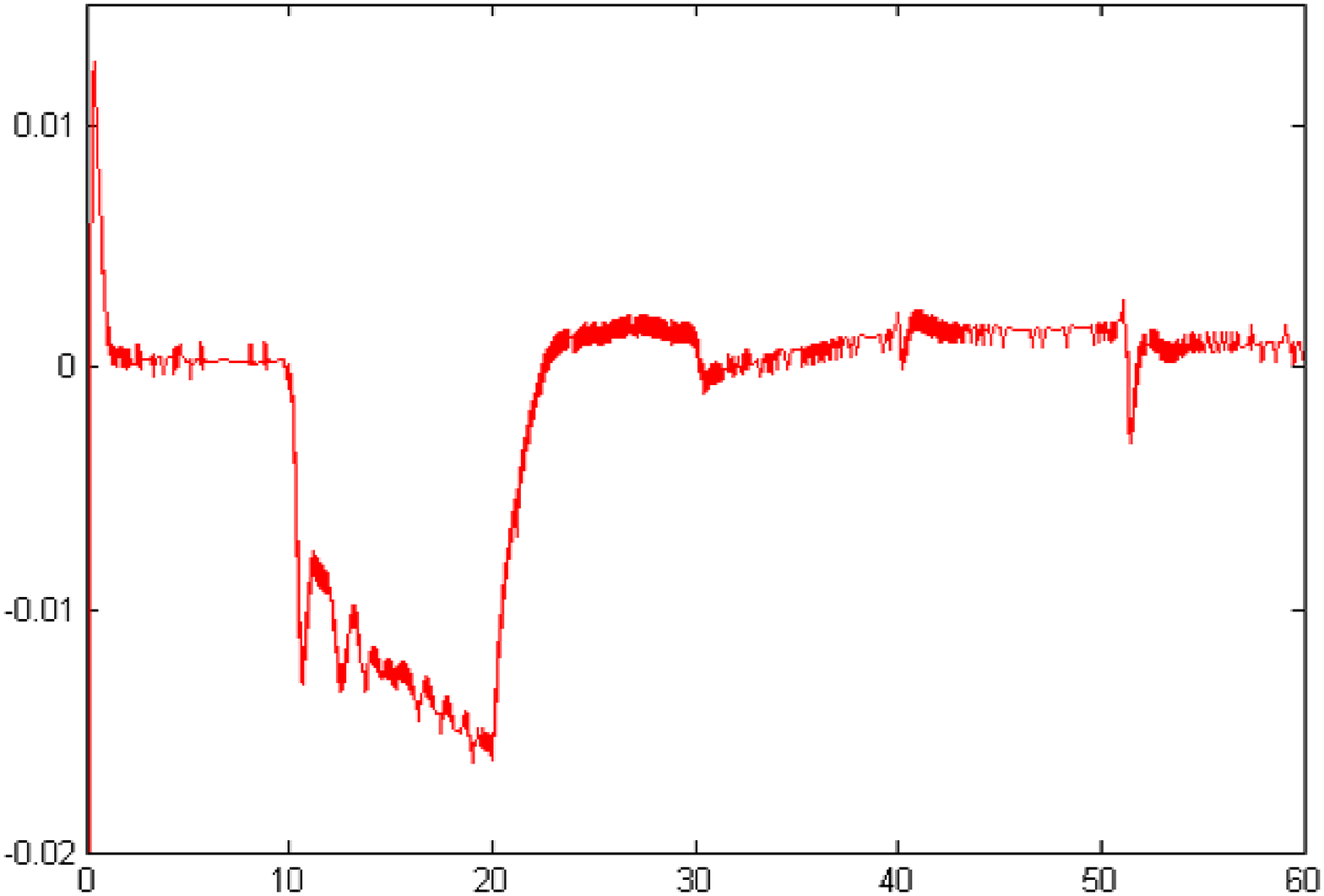}
\end{center}
\caption{UIO method: first residual detects the first fault}
\label{Pic_UIOr1}
\end{figure}

\begin{figure}[htbp]
\begin{center}
\includegraphics
[scale=0.23]{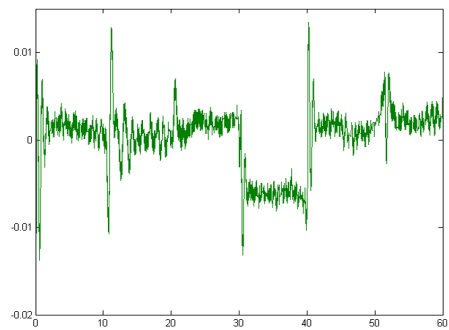}
\end{center}
\caption{UIO method: second residual detects the second fault}
\label{Pic_UIOr2}
\end{figure}

\begin{figure}[htbp]
\begin{center}
\includegraphics
[scale=0.23]{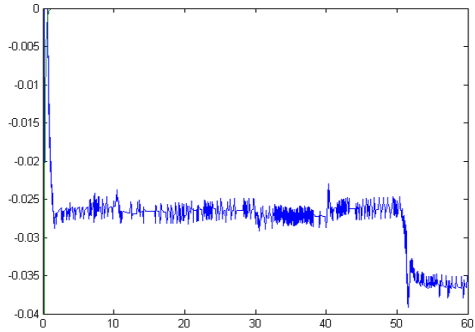}
\end{center}
\caption{UIO method: third residual detects the third fault}
\label{Pic_UIOr3}
\end{figure}

Figures \ref{Pic_UIOr1}--\ref{Pic_UIOr3} depict the residual signals of three detectors designed using the second method, UIO, to detect faults on travel angle, peach angle and elevation angle, respectively. Three different detectors are designed using UIO method. In each design, only one of the faults is supposed to be detected and the others are bypassed. It can be seen that all the three faults are detected while each detector is sensitive only to one of the faults. Hence, fault isolation requirements are satisfied. From the figures, it seems that the first method provides isolation in some extent while not as good as the second method. However the acceptable result is obtained by trial and error while the second method is more systematic for this purpose.

\begin{figure}[htbp]
\begin{center}
\includegraphics
[scale=0.23]{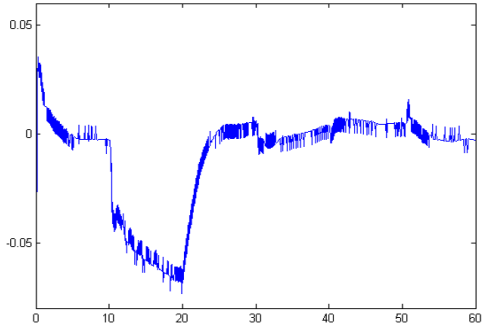}
\end{center}
\caption{Fault estimation using sliding-mode: first residual detects the fist fault}
\label{Pic_Sr1}
\end{figure}

\begin{figure}[htbp]
\begin{center}
\includegraphics
[scale=0.23]{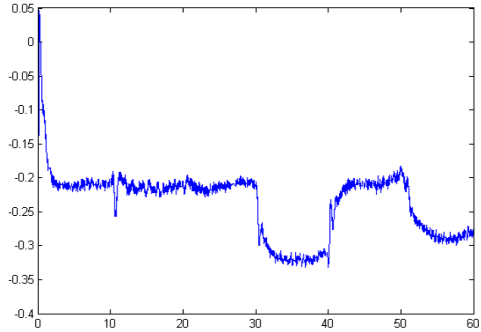}
\end{center}
\caption{Fault estimation using sliding-mode: second residual detects the second and the third fault}
\label{Pic_Sr2}
\end{figure}

Estimates of the faults obtained by sliding-mode detector are presented in Figures \ref{Pic_Sr1}--\ref{Pic_Sr2}. One signal shows the travel angle changes while the other signal shows the changes of the other two angle.

In the second experiment, we try to detect loss of control during flight \cite{wu2000detection}. One reason that such faults  may take place is that the performance of the actuators degrades. This fault changes the behavior of the system and the steady state of the system. Note that we do not design new observer detectors, but try to detect the loss of control effects using the observers designed for the first experiment. As the steady state values change due to this fault, we expect it to be detected by some of those observers. To model the loss of control effects, we send the control command $\bar{u}(t)$ to the system in place of $u(t)$ where \cite{wu2000detection}
\begin{eqnarray*}
\bar{u}(t)= X u(t).
\end{eqnarray*}
The second experiment takes 30 seconds. In nominal mode $X=I$. After 10 seconds loss of control occurs and $X$ changes to
\begin{eqnarray*}
X=\begin{pmatrix} 0.95 & 0\\ 0 & 0.3 \end{pmatrix}.
\end{eqnarray*}
For the rest of the experiment $X$ does not change.

\begin{figure}[htbp]
\begin{center}
\includegraphics
[scale=0.23]{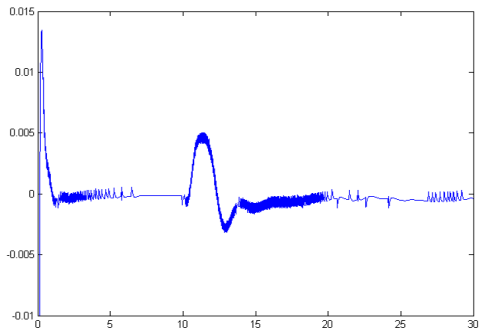}
\end{center}
\caption{Output observer method: first residual}
\label{Pic_r1_act}
\end{figure}

\begin{figure}[htbp]
\begin{center}
\includegraphics
[scale=0.23]{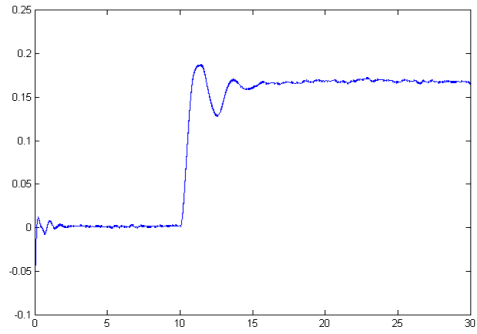}
\end{center}
\caption{Output observer method: second residual}
\label{Pic_r2_act}
\end{figure}

\begin{figure}[htbp]
\begin{center}
\includegraphics
[scale=0.23]{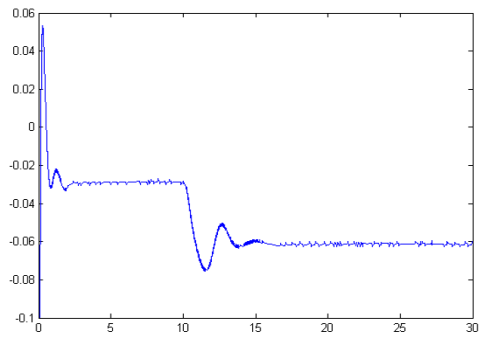}
\end{center}
\caption{Output observer method: third residual}
\label{Pic_r3_act}
\end{figure}

Again, experiment is repeated using the three fault detection methods. Figures \ref{Pic_r1_act}--\ref{Pic_r3_act} show the three residuals produced by the output observer method. The second and the third residuals clearly show a change while the effect of the change in the third residual disappears in steady state.

\begin{figure}[htbp]
\begin{center}
\includegraphics
[scale=0.23]{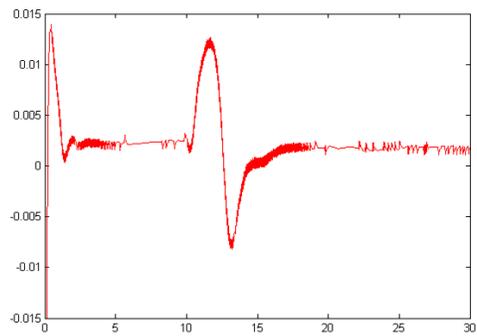}
\end{center}
\caption{UIO method: first residual}
\label{Pic_UIOr1_act}
\end{figure}

\begin{figure}[htbp]
\begin{center}
\includegraphics
[scale=0.23]{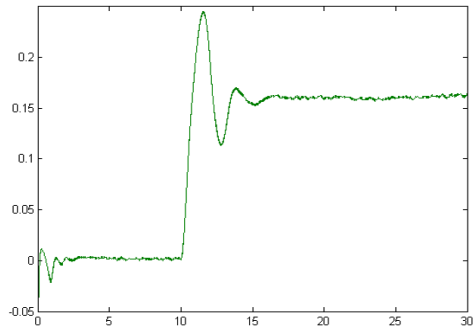}
\end{center}
\caption{UIO method: second residual}
\label{Pic_UIOr2_act}
\end{figure}

\begin{figure}[htbp]
\begin{center}
\includegraphics
[scale=0.23]{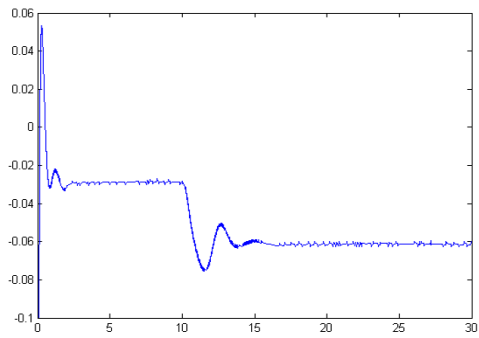}
\end{center}
\caption{UIO method: third residual}
\label{Pic_UIOr3_act}
\end{figure}

Figures \ref{Pic_UIOr1_act}--\ref{Pic_UIOr3_act} present the residuals of three detectors designed using UIO method. Two residuals out of three show a change in their steady state values.

\begin{figure}[htbp]
\begin{center}
\includegraphics
[scale=0.23]{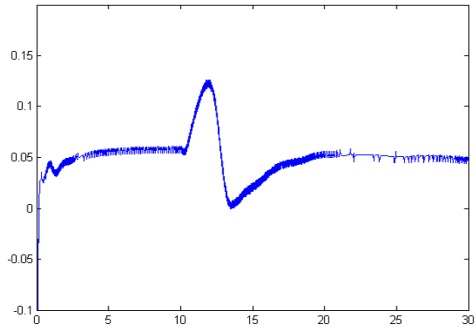}
\end{center}
\caption{Fault estimation using sliding-mode, first residual, steady state change is less than transient change in the residual}
\label{Pic_Sr1_act}
\end{figure}

\begin{figure}[htbp]
\begin{center}
\includegraphics
[scale=0.23]{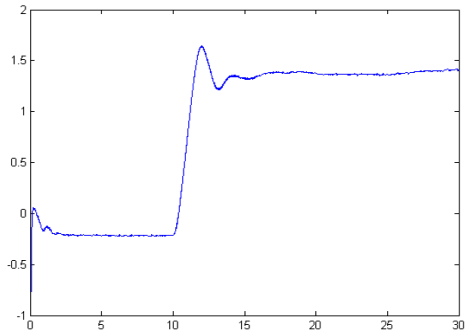}
\end{center}
\caption{Fault estimation using sliding-mode: second residual clearly show the occurrence of the fault}
\label{Pic_Sr2_act}
\end{figure}

The fault can be detected by using sliding-mode detector and the results are presented in Figures \ref{Pic_Sr1_act}--\ref{Pic_Sr2_act}.

We remind that chattering effect, which is a serious problem in sliding-mode control, is not very critical in fault detection. While the chattering in control may cause problems for actuators, the residual in fault detection is merely used to decide on the occurrence of the faults.

\begin{rem}
In this experiment, the real input $u(t)$ is provided to the detector while $\bar{u}(t)$ is sent to the system. The matrix $X$ models an actuator fault degradation which is supposed to be inside the system. However, the controller calculates the control signal correctly and provides it to the detector.
\end{rem}

\section{Fault detection software verification: towards formal methods} \label{veri}

The theories supporting three fault detection methods are presented in Section \ref{theory}. The methods are successfully applied to a real system in Section \ref{exper}. However, those methods are implemented as softwares and there always exist a semantic gap between theory and real software implementation. The methods in Section \ref{theory} should be implemented, either in graphical programming language tools such as Simulink or as computer codes in computer languages such as Matlab or C. Due to the different real life restrictions and computation errors, there might be a difference between implemented method in software form and the ideal method in theory. Simulations and practical tests are used to show that the methods work well. However, we cannot test the system for all possible conditions, initial values and faults. For that reason, we are interested in software verification approaches to include original design properties in the software and verify the correct operation of the software for all possible scenarios, based on the theory, and at least based on the model of the system used in the original design.
We aim at annotating the software so that an expert or a machine can track the operation of the software and verify that the original design criteria are satisfied at software level. The idea developed in \cite{feron2010control} to formally document the stability of closed-loop systems is extended to fault detection methods in this section. Here, we focus on the fault detection methods introduced in Section \ref{theory}

\subsection{Output observer} \label{veri_output}

The first system specification we must document in the software is the stability of the detector, i.e the variables of the observer software stay bounded over time. We need to verify the stability of error dynamics  in \eqref{5}--\eqref{5a}. In addition, we need to check that the fault detection duty of the software is performed correctly.
More precisely we need to show that
\begin{enumerate}
  \item Observer stability is guaranteed: the error dynamics are stable, i.e. $e(t)$ in \eqref{5} stays around origin when system is in nominal mode and remains bounded in faulty mode.
  \item Fault detection performance predicted by theory is guaranteed: the residual, $r(t)$, correctly detects the fault. In other words the residual is not around zero if the fault signal (for this method including both $f(t)$ and $f_d(t)$) passes a predefined threshold. On the other hand $r(t)$ stays around zero if the fault does not reach the threshold.
\end{enumerate}

Here, we start from the informal method in Section \ref{output}. We translate its properties into a formal language, compatible with software verification tools. Here, we assume that the closed-loop system is
stable, and we have already verified the stability of the closed-loop system using the method in \cite{feron2010control}. The result is that $y(t)$, and u(t) accordingly,
remain bounded in \eqref{3}--\eqref{4}. Thus, we only verify that error dynamics, $e(t)$, is bounded. This assumption is necessary, as $\hat{x}(t)$ blows up if $y(t)$ goes to infinity while $e(t)$ may remain bounded.

One approach to examine the stability of dynamic systems is to use Lyapunov theory, which is shown, on the other hand, to be a very efficient tool  in control software verification (see \cite{feron2010control}). We can check the stability by calculating the place of system poles if the system is linear, as well. However, it is preferred for software verification purposes to express desired properties in the form of invariant sets to which software variables must belong. We remind that in software implementation of the fault detection filter, observer state vector $\hat{x}(t)$ is a software variable (or a set of variables). Also, having the model of the system, a piece of code can represent the operation of the nominal system, with $x(t)$ a software variable. This code is synchronised with the detection filter (see closed-loop verification in \cite{feron2010control}).

\subsubsection{Steady state properties} \label{output-steady}
According to the properties above, there are two modes of the detection filter with respect to the observed system that we are mainly interested in to verify:
\begin{enumerate}
\item The detector is stable ($e(t)$ is bounded), the system is in nominal mode (fault signal is small enough, in terms of norm) and it is correctly detected ($r(t)$ is smaller than the predefined threshold).
\item The detector is stable ($e(t)$ is bounded), the system is faulty mode (fault signal is big enough) and it is correctly detected ($r(t)$ is larger than the predefined threshold).
\end{enumerate}
Both these properties are related to steady state specifications of the system. Other specifications might be important for the designer that are related to transient specifications. We discuss them in Section \ref{further}.

Consider a Lyapunov function $V(t)=e^T(t)Pe(t)$ with a proper positive definite $P$. We know that $V(t)$ decreases along system state trajectories and $\dot{V}(t)<0$. The following ellipsoid set is the invariant set in which $e(t)$ resides if the error dynamics are stable
\begin{eqnarray}
\mathcal{E}_n=\{e(t)\in \Re^n | e^T(t)Pe(t) \leq \zeta \},\label{45a}
\end{eqnarray}
for $\forall t \in \Re$, where $\zeta\geq0$ is a scalar. $\zeta$ is related to the threshold on $r(t)$. There are different ways to define a threshold on the residual. We may bound the peak value, the norm, the average norm over a time interval, etc. of $r(t)$ (see \cite{ding2008model}).
Consequently, fault alarm is raised if $\|r(t)\|_{\_}>r_{th}$ where $\| . \|_{\_}$ can be any norm, and $r_{th}$ is the threshold.
In addition to the common norms in robust control theory, we are interested in the peak amplitude of the residual vector. Also in\eqref{45a} we deal with peak value. We use the peak-norm which has been used in fault detection studies \cite{ding2008model}
\begin{eqnarray}
\|r\|_{peak}=\sup_t (r^T(t)r(t))^{\frac{1}{2}}.\label{63}
\end{eqnarray}
 Notice that $r(t)=Ce(t)$, namely we compare only the sensor measurements (and not the other quantities that are not directly measured) with the estimated values by the observer. We need to translate this threshold in the state space, define the reachable set of system trajectories corresponding to the threshold, and select $\zeta$. To this end, we need to find the system norm of the detector, $\pi=sup_{\bar{f}(t)\neq 0}\|r(t)\|_{\_}/\|\bar{f}(t)\|_{\_}$ where $\bar{f}(t)$ is defined in \eqref{60}. This can be calculated using LMI's, and depends on the norm that has been used to define the threshold.
Using $\pi$ and $r_{th}$ we can find the maximum norm of the augmented fault signal $\bar{f}(t)$ that is allowed by the initial design so that an alarm is not raised. So far, we know the set of fault signals that are tolerable by the detection observer (do not raise an alarm). The next step is to find $\bar{\pi}=sup_{\bar{f}(t)\neq 0}\|P^{\frac{1}{2}}e(t)\|_{2} / \|\bar{f}(t)\|_{\_}$ solving some LMI's. Finally $\zeta$ is obtained knowing $\bar{\pi}$ and the maximum norm of $\bar{f}(t)$ that does not raise an alarm.

For the faulty mode, we can introduce another ellipsoid invariant set around the new equilibrium point. However, we do not know where the new equilibrium point is, as the fault signals are assumed to be completely unknown. But in practice, $f(t)$ and $f_d(t)$ are norm bounded. Suppose that $\|\bar{f}(t)\|<\bar{\sigma}$. We introduce the invariant set below
\begin{eqnarray}
\mathcal{E}_f=\{e(t)\in \Re^n | e^T(t)Pe(t) \leq \bar{\zeta}\}.\label{46a}
\end{eqnarray}
In \eqref{46a} $\bar{\zeta}$ is
\begin{eqnarray}
\bar{\zeta}(t)&=& \max_{e(t),e(0) \in \mathcal{E}_n} e^T(t)Pe(t) \nonumber\\
& & s.t. \;\;\; \dot{e}(t)=(A-LC)e(t)+\bar{E}_f \bar{f}(t) \nonumber\\
& & \text{and} \;\;\; \|\bar{f}(t)\|_{\_}<\bar{\sigma}.\label{47a}
\end{eqnarray}

Hence, we have introduced two ellipsoid sets. As far as $V(t) \in \mathcal{E}_n$ for $\forall t \in \Re$, i.e. $\mathcal{E}_n$ is an invariant set, the system is in nominal mode and the detector is stable. Whenever $V(t) \in \mathcal{E}_f$ for $\forall t \in \Re$ so that $\exists t, V(t) \not\in \mathcal{E}_n$, i.e. $\mathcal{E}_f$ be an invariant set while $\mathcal{E}_n$ is not, the system is in faulty state and the detector is stable. We include all this information in the software as annotations. Other engineers who test the code can understand how the state space variables are supposed to behave in terms of invariant sets, without verifying the place of poles directly. If the annotations are expressed in a standard format, software verification tools can check it and certify that the software works well according to the original design.

To conclude this section we propose following Lemmas, which helps to calculate $\zeta$ and $\bar{\zeta}$. More results can be found in \cite{haddad2008nonlinear, chellaboina2000induced}

\begin{lemma} \label{lm1}
given the system
\begin{eqnarray}
\dot{x}(t)=Ax(t) + E d(t), x(0)=0, \label{48}
\end{eqnarray}
for a given constant $\rho_1>0$ where $P_1$ is a positive definite symmetric matrix we have
\begin{eqnarray}
\sup_t (x^T(t)P_1x(t))^{\frac{1}{2}}<\rho_1 \|d(t)\|_2, \label{49}
\end{eqnarray}
if there exist $Q_1>0$ so that
\begin{eqnarray}
\begin{pmatrix}
A^TQ_1+Q_1A &Q_1 E \\
E^TQ_1 &  -\rho I
\end{pmatrix}<0,\;\;
\begin{pmatrix}
Q_1 & P_1^{\frac{1}{2}} \\
 P_1^{\frac{1}{2}} &  \rho_1 I
\end{pmatrix}>0. \label{50}
\end{eqnarray}
Also if $y(t)=Cx(t)$ then for a given $\bar{\rho}_1>0$
\begin{eqnarray}
\|y\|_{peak}<\bar{\rho}_1 \|d(t)\|_2, \label{49a}
\end{eqnarray}
if there exist $\bar{Q}_1>0$ so that
\begin{eqnarray}
\begin{pmatrix}
A^T\bar{Q}_1+\bar{Q}_1A &\bar{Q}_1 E \\
E^T\bar{Q}_1 &  -\bar{\rho}_1 I
\end{pmatrix}<0,\;\;
\begin{pmatrix}
\bar{Q}_1 & C^T \\
C &  \bar{\rho}_1 I
\end{pmatrix}>0. \label{50a}
\end{eqnarray}
\end{lemma}

\begin{lemma} \label{lm2}
given the system
\begin{eqnarray}
\dot{x}(t)=Ax(t) + E d(t), y(t)=Cx(t), x(0)=0, \label{48b}
\end{eqnarray}
where
\begin{eqnarray}
\forall t, \;\; d^T(t) d(t)<0, \label{65}
\end{eqnarray}
for a given constant $\rho_2>0$ where $P_2$ is a positive definite symmetric matrix we have
\begin{eqnarray}
\sup_t (x^T(t)P_2x(t))^{\frac{1}{2}}<\rho_2 \|d(t)\|_{peak}, \label{49b}
\end{eqnarray}
if there exist scalars $\upsilon>0$ $\varphi>0$ and a matrix $Q_2>0$ so that
\begin{eqnarray}
\begin{pmatrix}
A^TQ_2+Q_2A+ \upsilon Q_2 &Q_2 E \\
E^TQ_2 &  - \varphi I
\end{pmatrix}<0,\;\; \nonumber \\
\begin{pmatrix}
\upsilon Q_2 & 0& P_2^{\frac{1}{2}} \\
0 &(\rho_2-\varphi) I & 0\\
 P^{\frac{1}{2}}_2 & 0 &  \rho I
\end{pmatrix}>0. \label{50b}
\end{eqnarray}
Also for a given $\bar{\rho}_2>0$
\begin{eqnarray}
\|y\|_{peak}<\bar{\rho}_2 \|d(t)\|_{peak}, \label{49ab}
\end{eqnarray}
if there exist $\bar{\upsilon}>0$ $\bar{\varphi}>0$ and $\bar{Q}_2>0$ so that
\begin{eqnarray}
\begin{pmatrix}
A^T\bar{Q}_2+\bar{Q}_2A +\bar{\upsilon} \bar{Q}_2&\bar{Q}_2 E \\
E^T\bar{Q}_2 &  - \bar{\varphi} I
\end{pmatrix}<0,\;\;\nonumber\\
\begin{pmatrix}
\bar{\upsilon} \bar{Q}_2 & 0& C^T \\
0 &(\bar{\rho}_2-\bar{\varphi}) I & 0\\
C & 0 &  \bar{\rho}_2 I
\end{pmatrix}>0. \label{50ab}
\end{eqnarray}
\end{lemma}

\begin{lemma} \label{lm3}
given the system
\begin{eqnarray}
\dot{x}(t)=Ax(t) + E d(t),\;\;y(t)=Cx(t),\;\; x(0)=0, \label{48c}
\end{eqnarray}
for a given constant $\rho_3>0$ we have
\begin{eqnarray}
\|y(t)\|_2<\rho_3 \|d(t)\|_2, \label{49c}
\end{eqnarray}
if there exist $Q_3>0$ so that
\begin{eqnarray}
\begin{pmatrix}
A^TQ_3+Q_3A &Q_3 E & C^T\\
E^TQ_3 & -\rho_3 I &0\\
C & 0 & -\rho_3 I
\end{pmatrix}<0, \label{50c}
\end{eqnarray}
\end{lemma}

\subsubsection{Instability of the detector} \label{unstable}

For the verification of commercial software, the properties expressed in Section \ref{output-steady} are certified.
It is obvious that the final software product guarantees the stability and correct performance.
There are other modes that are undesired, however. We specify those modes to help the test engineers to detect errors of the software, before finalizing the product.
Based on Lyapunov theory for linear systems, the detector is unstable if $\mathcal{E}_f$ is not invariant.

\subsubsection{Further analysis and transient mode} \label{further}

Verifying that $\mathcal{E}_n$ or $\mathcal{E}_f$ is invariant ensures the steady state properties of the detection filter, as we discussed in Section \ref{output-steady}. However, transient mode, namely changing from nominal to faulty mode, cannot be completely analysed. In this section, other properties obtained from Lyapunov theory and the theory of linear systems control and observation are used to complete the documentation of the software. We aim at verifying the mode of the system at each time instant. For instance if $V(t) \in \mathcal{E}_f - \mathcal{E}_n$ at time $t$, we cannot infer whether the system is faulty or the detector is unstable using the material in in Section \ref{output-steady}. We need to prove that
$\forall t, V(t) \in \mathcal{E}_f$ to conclude the system is faulty. In this section, we aim at concluding some results on any state trajectory point that software variables reach, before investigating the entire set.
Let us introduce the following new variable
\begin{eqnarray}
\theta(t)= \dot{e}(t)-(A-LC)e(t),\label{58}
\end{eqnarray}
According to the original design requirements, the system is in nominal mode if and only if $\|\theta(t)\|_{\_}<\theta_{th}$. It is faulty if and only if $\|\theta(t)\|_{\_}\geq \theta_{th}$. Now suppose that system trajectory is in $\mathcal{E}_f - \mathcal{E}_n$ and the software verification tool is trying to verify whether or not $\mathcal{E}_f$ is invariant. Assume, $\|\theta(t)\|_{\_}<\theta_{th}$; it means the system is in nominal mode, and therefore $V(t) \in \mathcal{E}_f - \mathcal{E}_n$ for some $t$ show that the detector is unstable and $\mathcal{E}_f$ cannot be invariant. The verification tool should be able to find $V(t+\tau) \not\in \mathcal{E}_f$ for some $\tau$ continuing the same trajectory. If such an example does not exist, there is an error in the code. Note $\theta_{th}$ depends on the threshold on fault signal. If we know the fault signal threshold
on each fault input channel, the exact threshold can be calculated. If the norm of the fault vector is available, un upper bound can be obtained from the triangle inequality.

Further results can be deduced from the derivative of the Lyapunov function. $\dot{V}(t)$ can be used to analyse the correct operation of the detector in some modes.
For instance, if $V(t) \in \mathcal{E}_n$ and and $\|\theta\|>\theta_{th}$, then we must have $\|\dot{V}(t)\|>\epsilon_{V}$ for some positive $\epsilon_{V}$. Otherwise some implementation errors exist in the code. A summary of all possible modes is provided in Table \ref{tab1}. Here, ``convergence issue'' refers to the instability of the observer.
In this table we suppose that faulty system will not return to nominal mode, as it is usually assumed. Removing this assumption leads to a more complected table and the derivative of the Lyapunov function should be used to distinguish between ``returning to nominal mode from faulty mode'' and the ``convergence issue''.
\begin{table}[h]
\centering
\caption{Verification table}
\begin{tabular}{| c | c | c | c |} %
 \hline
 & $V(t) \in \mathcal{E}_n$ & $V(t) \not\in \mathcal{E}_n$, & $V(t) \not\in \mathcal{E}_f$  \\
 & & $V(t) \in \mathcal{E}_f$ &  \\
\hline
$\|\theta\|<\theta_{th}$  & Nominal  & Convergence  & Convergence   \\
                       & mode    & issue & issue   \\
\hline
$\|\theta\|>\theta_{th}$  & Transient mode & Faulty & Convergence  \\
                       & if $\|\dot{V}(t)\|>\epsilon_{V}$ & mode & issue\\
\hline
\end{tabular}\label{tab1}
\end{table}

\begin{figure}[htbp]
\begin{center}
\includegraphics
[scale=0.40,angle=0]{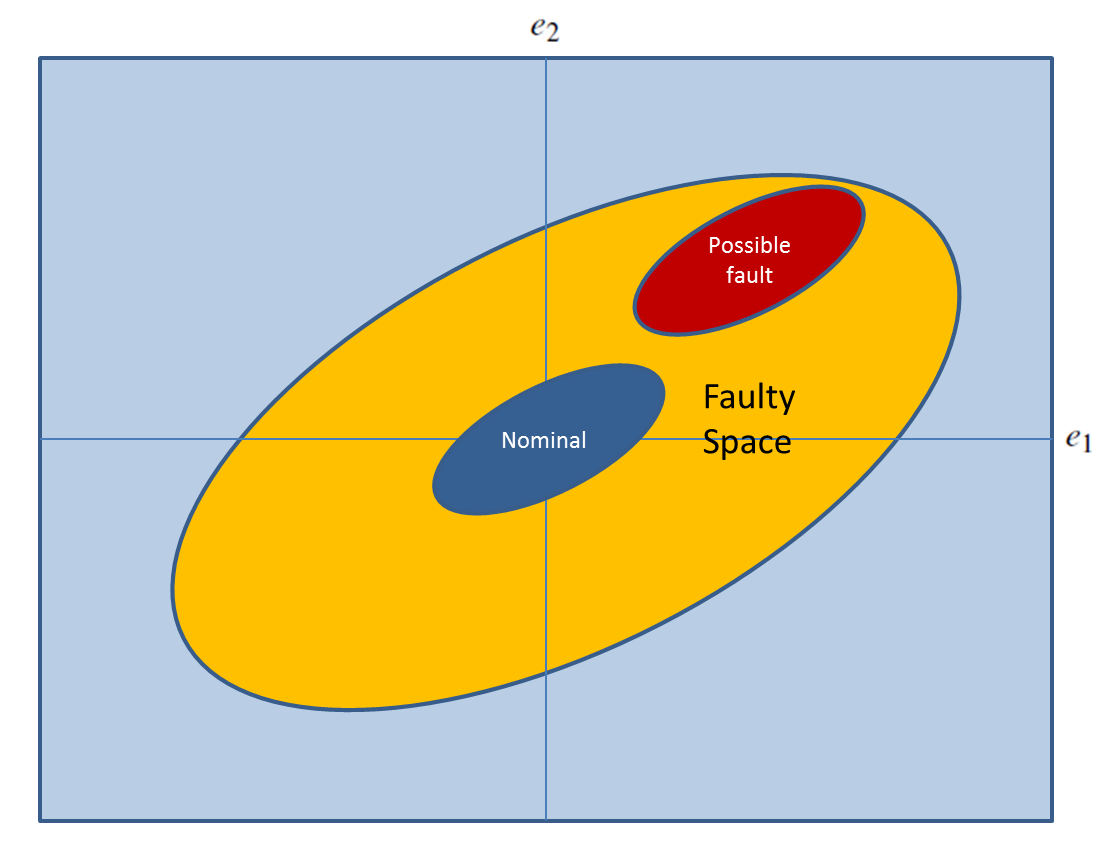}
\end{center}
\caption{Blue ellipsoid shows $\mathcal{E}_n$ and yellow ellipsoid demonstrates $\mathcal{E}_f$ in error space. Red ellipsoid is one possible fault scenario.}
\label{Lyapunov}
\end{figure}

Figure \ref{Lyapunov} shows the space of error and the nominal and faulty regions. The order of the system is two, in order to be able to draw the error space in two dimensions.

\begin{figure}[htbp]
\begin{center}
\includegraphics
[scale=0.40,angle=0]{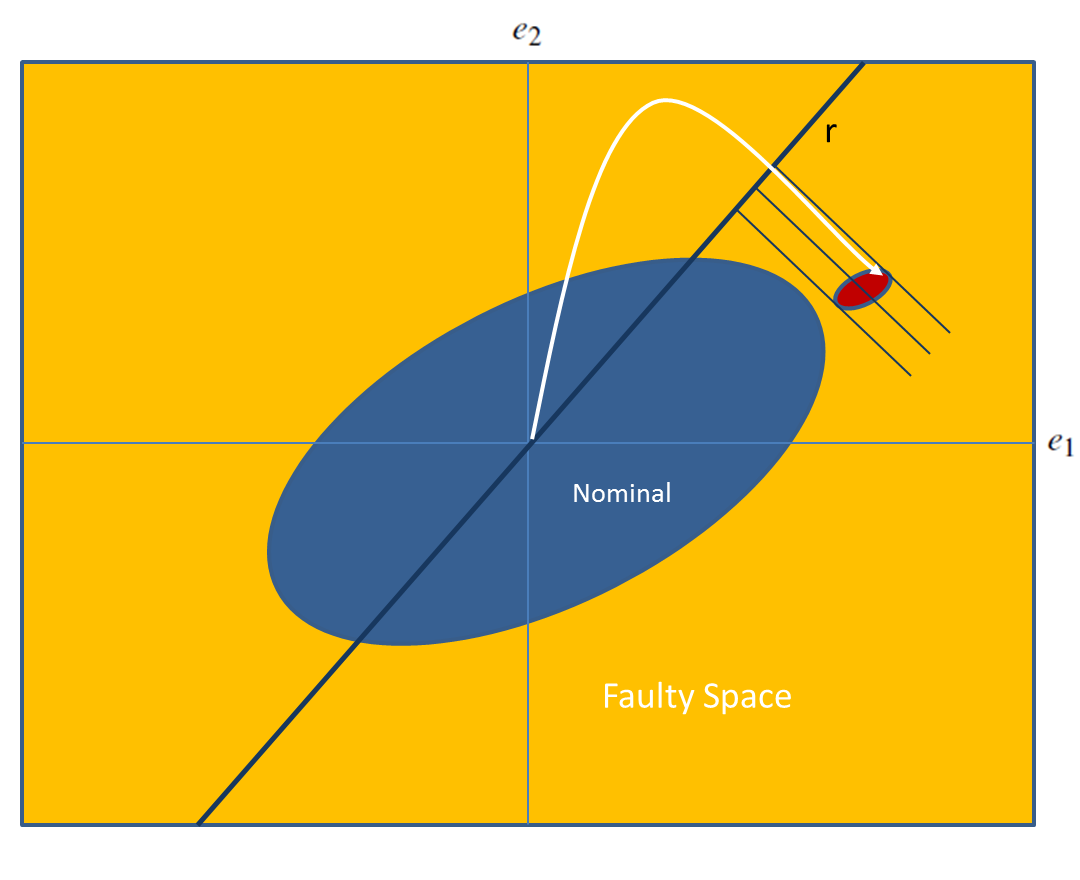}
\end{center}
\caption{Unit step is applied to the one input. After Transient time $t_s$, $\mathcal{E}_t$ is invariant.}
\label{Transient}
\end{figure}

One property that is important in transient response of linear systems is settling time $t_s$. In safety-critical systems, it is of great importance to verify that the fault detection mechanism works fast enough to detect fault before they lead to catastrophic consequences. Settling time is the time elapsed from the application of a step input to the time at which the  output has entered and remained within a specified error band, usually symmetrical about the final value (see \cite{ogata1970modern}). The same concept can be introduced and be verified for the detection observers. We are interested in testing the settling time to verify that the observer is fast enough, according to the original design. The settling time can be computed as described in \cite{ogata1970modern}. Here, we translate the idea in the space of errors and based on invariant sets for verification purposes. As such, it will be possible to certify that property in the code and track possible changes in settling time due to the modifications that can be possibly made by each line of the code. Settling time depends on the place of the observer poles. For each pair of input-residual, we can find the corresponding settling time and the steady state value of the error, $e_s$. Applying such a unit step input as fault signal, the following set must become invariant after $t_s$ seconds.
\begin{eqnarray}
\mathcal{E}_t=\{e(t)\in \Re^n | (e(t)-e_s)^TP(e(t)-e_s) \leq \zeta_t \},\label{64}
\end{eqnarray}
Here, $\zeta_t$ is selected so that the projection of $\mathcal{E}_t$ on the corresponding residual direction is equal to the error bound around the steady state of the residual after settling time. Figure \ref{Transient} shows how to verify the settling time. Note that the settling time can be calculated from the place of dominant poles of the detection filter. On the other hand, we can express these transient time response properties in the form of LMI's, \cite{boyd1997linear,scherer1997multiobjective}. An easier way to obtain an upper bound on settling time for multi input multi output systems is to calculate the settling time from an input to all states and select the maximum.

\begin{rem}
Derivative calculation is not straightforward. Particular high pass filters of the form
\begin{eqnarray}
T(s)= \frac{s}{1+a s}.\label{61}
\end{eqnarray}
should be implemented to approximate the derivative, where $a$ is a small scaler and $s$ is Laplace transform variable.
\end{rem}

It is important to know that software verification is an approach to check the mathematical proofs in the software level. If some properties of a dynamic system seems to be correct in some cases in simulation or practical tests but concrete proofs does not support the general case, software verification cannot automatically prove it. The method of output observer implemented for fault detection uses a general purpose observer to detect additive faults and does not claim on fault isolation. Although one may show isolation properties by simulations or practical experience, as we do in Section \ref{exper} for our particular application and fault scenarios, the method originally does not provide proofs. Consequently, fault isolation is not verified for this method. On contrary, UIO provides fault isolation properties, as we discuss in Section \ref{veri_UIO}. Therefore, the proof can be implemented for software verification. This is an important advantage of using UIO, even though output observers give good results in practice.

\subsubsection{Annotation of the software implemented on the helicopter}

Here we consider the last fault scenario described in Section \ref{scenario}. The following $P$ is used for the verification purpose
\begin{eqnarray*}
P=\begin{pmatrix}0.3306 & 0 &  0  & -0.0730  &  0  &  0\\
0 & 0.2937 &  0  &  0  & -0.0799 & 0\\
0 & 0 & 0.2908 & 0 &  0 & -0.0802\\
-0.0730 & 0& 0 & 0.0221 & 0  & 0\\
0 & -0.0799  & 0 & 0 & 0.0312 & 0\\
0 & 0 & -0.0802 & 0 & 0 & 0.0317
\end{pmatrix}
\end{eqnarray*}
Assume the threshold on the 2-norm of the residual is selected as $r_{th}$ by the designer. Using Lemma \ref{lm3} we can compute the $H_{\infty}$ norm of the observer. To compute the norm we need to solve LMI's by YALMIP \cite{lofberg2004yalmip}. Hence the maximum $\rho_3$ that satisfies \eqref{49c} is $0.0086$ and the corresponding $Q_3$ obtains as bellow
\begin{eqnarray*}
\!\!Q_3\!\!\!\!=\!\!\!\!\begin{pmatrix}
43.0392 & 3.3479 & 0.0001 & -8.9977 & -0.9984  &  0\\
3.3479 &  7.2098 &  0 & -0.7865 & -0.5913 & 0\\
0.0001 & 0  & 50.9619 & -0.0000  & 0 & -13.5786\\
-8.9977 &  -0.7865 &  0 &  2.6560 & 0.2945 & 0\\
-0.9984  & -0.5913 & 0 & 0.2945 & 0.2473 & 0\\
0 & 0 & -13.5786 & 0 & 0  & 5.2048
\end{pmatrix}
\end{eqnarray*}
A 2-norm bound on the fault signals that are not suppose to raise a fault alarm is $116.0542 r_{th}$. Using Lemma \ref{lm1}, an upper bound on $\sup_t (x^T(t)P_1x(t))^{\frac{1}{2}}$ is $3.0783 r_{th}$ and the corresponding $Q_1$ is
\begin{eqnarray*}
\!\!Q_1\!\!\!\!=\!\!\!\!\begin{pmatrix}
2167.7 & 307.9 & 0 & -479.7 & -66.2 & 0\\
307.9 & 1114.8 & 0 & -106.7 & -76.5 & 0\\
0 & 0 & 2594.8 & 0 & 0 & -742.4\\
-479.7  & -106.7 & 0 & 145.2 & 21.8 & 0\\
-66.2 & -76.5 & 0 & 21.8 & 7.7 & 0\\
0 & 0 & -742.4 & 0 & 0 & 289.5
\end{pmatrix}
\end{eqnarray*}
Accordingly, we select $\zeta=9.4761*r_{th}^2$. If $r_{th}=.2$ we obtain $\zeta=0.3790$. Assume the maximum norm of possible fault signal is $3$ times
bigger than the minimum fault signal that triggers a fault alarm. Therefore $\bar{\zeta}=3.4114$.
Another parameter we need to calculate is $\theta_{th}$. An upper bound on $\|\theta\|$ is $9.5647$.
The derivative of the Lyapunov function is $e^T(t) Q e(t)$ where
\begin{eqnarray*}
\!\!Q\!\!\!\!=\!\!\!\!\begin{pmatrix}
0 & 0 & 0 & 0.3306 & 0 & 0\\
0 & 0 & 0.0987 & 0 & 0.2937 & -0.0390\\
0 & 0.0987 & 0 & 0 & 0 & 0.2908\\
0.3306 & 0 & 0 & -0.1460 &  0 & 0\\
0 & 0.2937 & 0 & 0 & -0.1598 & 0\\
0 & -0.0390 & 0.2908 & 0 & 0 & -0.1605
\end{pmatrix}
\end{eqnarray*}
The upper bound on settling time for this observer is $0.8479 sec$.

\subsection{Unknown input observer} \label{veri_UIO}

The stability of the fault detection filter is the first property, again, that we need to document in the software in a formal manner.
Also, we need to show that the software is capable of detecting unwanted inputs.
However, fault isolation should be consider in addition to the fault detection property, unlike for output observer detectors in Section \ref{veri_output}.
Here, we summarize the main properties we need to be verified:
\begin{enumerate}
  \item Stability is guaranteed: the dynamics of the error remains bounded whether the system is in nominal mode or in faulty mode.
  \item Fault detection performance is guaranteed: the detector works correctly, i.e. the residual $r(t)$ is around zero if fault $f(t)$ is around zero and deviates from zero if $f(t)$ is greater than a predefined threshold.
  \item Fault isolation is certified: the faults that we do not want to detect using the filter, $f_d(t)$, have no (or negligible) influence on $r(t)$.
\end{enumerate}
As in Section \ref{veri_output}, we assume that the closed-loop system stability is verified and $y(t)$ and $u(t)$ stay bounded.
 Again, we use Lyapunov theory to translate the stability property into the formal language. For the verification of fault detection performance, we need to show that a fault signal that is strong enough results in a fault alarm. These two parts are similar to the verification of output observer in Section \ref{veri_output}. We briefly discuss them, without going through the details. The last property, the verification of fault isolation, is however new and we discuss it in detail. In the software implementation of the detector, $z(t)$, $u(t)$, $y(t)$ are the variables of detection filter software. Also, the system model can be implemented as a piece of code with $x(t)$ a software variable. These two softwares are synchronized and work together.
\subsubsection{Fault isolability} \label{isolability}
We need to introduce a new variable, $\theta(t)$, to verify fault isolability. The same variable will be used for further analysis as in Section \ref{further}, later. To explain the reason, we need to return to the error dynamics for UIO. Consider equation \eqref{14} and the following inequalities
\begin{eqnarray}
& &\|[F-(A-HCA-K_1C)]z(t)\| \leq \epsilon,  \label{41a}\\
& &\|[K_2-(A-HCA-K_1C)H]y(t)\| \leq \epsilon,  \label{41b}\\
& &\|[T-(I-HC)]Bu(t)\| \leq \epsilon, \label{41c}
\end{eqnarray}
where $\epsilon \geq 0$ is a small scalar. According to the discussions in Section \ref{UIO}, $\epsilon$ must be equal to zero for an error-free system. However, software variables are floating numbers and not real numbers and the right hand of \eqref{15} is never exactly zero. Also computation errors always exist in the software implementation. Hence, we introduce an error tolerance for computational errors by $\epsilon$. These conditions are added to the code of the detector as annotations, before verifying the stability. If \eqref{41a}--\eqref{41c} are satisfied, only fault $f(t)$ and $f_d(t)$ are the major exogenous inputs to the error dynamics.
As the fault signals are unknown, we are not able to evaluate the those terms in \eqref{14}. However, it is possible to work with $\theta(t)$ defined bellow
\begin{eqnarray}
\theta(t)&=&\dot{e}(t)-(A-HCA-K_1C)e(t)\nonumber\\
&-&[F-(A-HCA-K_1C)]z(t) \nonumber \\
&-&[T-(I-HC)]Bu(t) \nonumber\\
&-&[K_2-(A-HCA-K_1C)H]y(t). \label{42}
\end{eqnarray}
Assume that \eqref{41a}--\eqref{41c} are satisfied. To cancel out the effect of $f(t)$ on $\theta(t)$, consider  $(E_fT)^{\perp}\theta(t)$, where $(E_fT)^{\perp}$ is the left annihilator of $E_fT$.
We need to verify that
\begin{eqnarray}
\|(E_fT)^{\perp}\theta(t)\| \leq \epsilon, \label{44}
\end{eqnarray}
through the code.
Here, \eqref{44} guarantees that the system is insensitive to $f_d(t)$. Note that \eqref{41a}--\eqref{41c} and \eqref{44} must be satisfied in all modes of the system and the detector, in order to guarantee the isolability. If we guarantee that the effect of $f_d(t)$ on the error dynamics is negligible, then the rest of verification process is similar to the verification of output observer. We briefly explain it bellow.
\subsubsection{Steady state properties}
In this section, we suppose that fault isolation conditions, \eqref{41a}--\eqref{41c} and \eqref{44}, are satisfied.
We need to document and verify two main operating modes
\begin{enumerate}
\item The detector is stable ($e(t)$ is bounded), the system is in nominal mode ($f(t)$ is small enough in terms of norm) and it is correctly detected ($r(t)$ is smaller than a predefined threshold). Isolability condition is satisfied ($r(t)$ is not affected by $f_d(t)$).
\item The detector is stable ($e(t)$ is bounded), the system is in faulty mode ($f(t)$ is not small enough) and it is correctly detected ($r(t)$ is larger than the predefined threshold). Isolability condition is satisfied ($r(t)$ is not affected by $f_d(t)$).
\end{enumerate}
 To verify the stability of the error dynamics we use Lyapunov theory, as in previous section. We suppose that the fault $f(t)$ acts as an step signal. But, the magnitude of the signal is unknown. It means that the equilibrium point of the system changes because of the fault. However, we do not have any information about the new equilibrium point. Considering a Lyapunov function $V(t)=e^T(t)Pe(t)$ for a positive definite $P$, we can show that the $e(t)$ remains in a predefined invariant ellipsoid
\begin{eqnarray}
\mathcal{E}_n=\{e(t)\in \Re^n | e^T(t)Pe(t) \leq \zeta \},\label{45}
\end{eqnarray}
for $\forall t \in \Re$, if the detector is stable and the system is in nominal mode. Here, $\zeta\geq0$ is a scalar.

Assuming that $\|f(t)\|<\sigma$. We introduce
\begin{eqnarray}
\mathcal{E}_f=\{e(t)\in \Re^n | e^T(t)Pe(t) \leq \bar{\zeta} \},\label{46}
\end{eqnarray}
for the faulty mode.
In \eqref{46} $\bar{\zeta}$ is
\begin{eqnarray}
\bar{\zeta}(t)&=& \max_{e(t),e(0) \in \mathcal{E}_n}  e^T(t)Pe(t) \nonumber\\
& & s.t. \;\;\; \dot{e}(t)=F e(t)-TE_f f(t) \nonumber\\
& & and \;\;\; \|f(t)\|<\sigma.\label{47}
\end{eqnarray}
If $\mathcal{E}_n$ is invariant, the system is in nominal mode and the observe is stable. Whenever $V(t) \in \mathcal{E}_f$ is invariant while  $\mathcal{E}_n$ is not, the system is faulty and the detector is stable. See the details given in Section \ref{output-steady}. Figure \ref{Lyapunov} is still valid for UIO, assuming that isolation requirements are satisfied.
\subsubsection{Instability of the detector}
If for some $t$ we have  $V(t) \not\in \mathcal{E}_f$ and the fault isolation conditions are satisfied, the dynamics of error and consequently the detector are unstable.

\subsubsection{Further analysis and transient mode}
Assume isolability conditions are satisfied. For further analysis we make use of $\theta(t)$ comparing its norm against a predefined threshold
\begin{eqnarray}
\|\theta(t)\|_{\_} \geq \theta_{th}, \label{43}
\end{eqnarray}
as in Section \ref{further}. We can also use $\dot{V}(t)$. The approach is the same and we do not repeat it. Transient mode is again specified by settling time.

\subsubsection{Annotation of the software implemented on the helicopter}
Even if all computations are correct inside the code, $(E_fT)^{\perp}$ is not usually zero due to the computation errors in null space.
In this example the order of the error is $10^{-15}$. This error is greater than any error in \eqref{41a}--\eqref{41c}. In order to be sure that
such an error will not propagate in the code and will not affect the isolability we select $\epsilon=10^{-10}$.
Calculating the other parameters is similar to those for the output observer and we do not give the details.

\subsection{Sliding-mode fault detector} \label{veri_sliding}

Sliding-mode observers are nonlinear. Some of nice properties that we can conclude for linear observers cannot be easily extended to sliding-mode observers. We discuss the properties that we can verify, in this section.

We need to document and verify the stability and correct estimation of the detector. Correct fault estimation needs that error dynamics reside on the sliding surface. For the sliding-mode observers, we need to verify the following properties
\begin{enumerate}
  \item The detector is stable ($e(t)$ is bounded),
  \item Sliding motion takes place on sliding surface and the error dynamics states remain on the sliding surface.
\end{enumerate}

We need to show that an sliding motion takes place on
\begin{eqnarray}
S_o=\{e\in \Re^n | Ce=0\},\label{52}
\end{eqnarray}
in finite time and the dynamics of error in \eqref{36}--\eqref{37} are stable. If these two conditions are satisfied, we can conclude \eqref{39} is correct, and consequently \eqref{40} is an estimation of the fault.
Based on the proofs in \cite{edwards1998sliding} the following invariant ellipsoids are considered for the verification of the sliding-mode method
\begin{eqnarray}
\mathcal{E}_s&=&\{e_y(t)\in \Re^p | e_y^T(t)P_2e_y(t) \leq \alpha \}, \label{53}\\
\mathcal{E}_e&=&\{e_1(t)\in \Re^{n-p} ,e_y(t)\in \Re^p | \nonumber\\
& & e_1^T(t)^TP_1e_1(t)+e_y^T(t)P_2e_y(t) \leq \beta \}, \label{54}
\end{eqnarray}
for sufficiently small predefined scalars $\alpha > 0$ and $\beta>0$, where
\begin{eqnarray}
& & P_2 A_{22}^s +(A_{22}^s)^T P_2=-Q_2,  \label{55}\\
& & \hat{Q}=A_{21}^TP_2Q_2^{-1}P_2A_{21}+Q_1,  \label{56}\\
& & P_1A_{11}+A_{11}^T P_1 =-\hat{Q},  \label{57}
\end{eqnarray}
for $Q_1 \in \Re^{(n-p) \times (n-p)}$ and $Q_2 \in \Re^{p \times p}$ be symmetric positive definite matrices.
We need the linear transformation $T_o$ to obtain $x_1(t)$ from $x(t)$. Details are given in \cite{edwards1998sliding}.

It has been shown that sliding motion takes place on sliding surface in finite time. We show that time interval by $t_s$. An upper bound on the time error dynamic need to reach the sliding surface is given in \cite{edwards1998sliding}.

\begin{figure}[htbp]
\begin{center}
\includegraphics
[scale=0.45,angle=0]{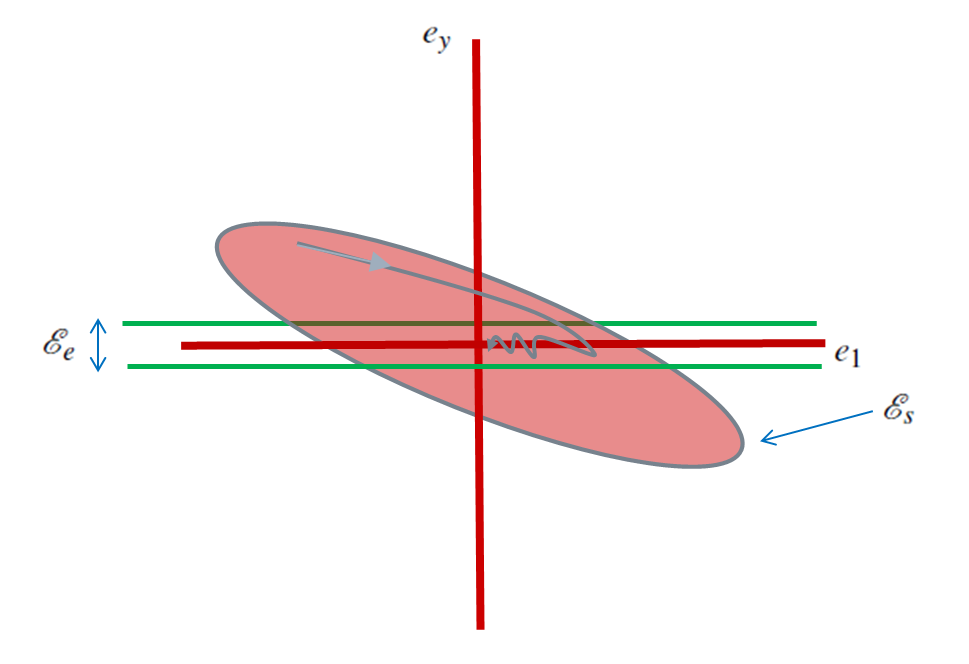}
\end{center}
\caption{Invariants sets defined for the verification of sliding-mode fault detection observer.}
\label{Sliding_fig}
\end{figure}

If $\mathcal{E}_e$ is an invariant set, and $\mathcal{E}_s$ is invariant after $t_s$, sliding motion takes place, dynamics of error is stable and consequently the detector can detect faults $\|\bar{f}(t)\| < \rho(t,y,u)$, according to the theory.
Figure \ref{Sliding_fig} shows the trajectories of dynamics of error and corresponding invariant sets.

Note that stability of the dynamics of error is necessary but not sufficient condition for correct fault detection. Sliding motion must take place as well. Hence, both $\mathcal{E}_e$ and $\mathcal{E}_s$ should remain invariant sets.

\subsubsection{Annotation of the software implemented on the helicopter}
Assume $Q_1$ is an identity matrix. The other parameters and Lyapunov matrices are obtained as bellow

\begin{eqnarray*}
Q_2&=&I_{3\times3},\\
P_1&=& \begin{pmatrix}
-0.5007 & 0 & 0\\
0 & -0.5011 & 0\\
0 & 0 & -0.0456
\end{pmatrix},\\
P_2&=&
\begin{pmatrix}
0.05 & 0 & 0\\
0 & 0.0417 & 0\\
0 & 0 & 0.0385\\
\end{pmatrix},\\
\hat{Q}&=&
\begin{pmatrix}
1.0015 & 0 & 0\\
0 & 1.0021 & -0.0004\\
0 & -0.0004 & 1.0021
\end{pmatrix}.
\end{eqnarray*}

\subsection{Discussions}

Three observers introduced in this paper are different from the view point of software documentation and verification. If the only purpose is the detection of any unknown input to the system, all three methods can be verified. However, if we consider fault isolation in the software level, the first method, output observer, cannot be verified, thought isolation might be achieved in particular cases in practice by trial and error.
In that regards, UIO and sliding-mode approaches are superior. Fault isolation and estimation properties can be verified as it is shown in Sections \ref{veri_output} and \ref{veri_sliding}.

Another point that should be considered is the possibility of extending the results to discrete-time systems. Although we implement methods in continuous-time and Simulink provides the possibility of implementing them to the helicopter model, all digital computers work in discrete-time in practice. As long as the sampling frequency is high, continuous-time behavior is simulated. However, it is sometimes proffered to design the detectors in discrete-time by sampling input-output data and using a discrete-time model of the system. As such, the performance will not depend on the sampling frequency. The theory of output observer design and UIO design can be easily extended to discrete-time systems. However, it is not straightforward to do the same for sliding-mode observers. To the best of our knowledge, the method in Section \ref{sliding} is not extended for discrete-time systems, yet. On the other hand tuning the parameter $\rho$ is not easy in practice when we design sliding-mode observers. That affects the correct estimation of the fault claimed by sliding-mode method. If this parameter is not selected correctly by the original designer, fault estimation reduces to isolation, i.e. the reconstructed fault estimate does not show the correct magnitude of the fault. Note the main advantage of sliding-mode method is to estimate the fault while the other two methods only claim to detect it. The linear structure of UIO, which is simpler than the nonlinear sliding-mode detector, is another advantage that should be consider if only fault isolation is required.
%%%%%%%%%%%%%%%%%%%%%%%%%%%%%%%%%%%%%%%%%%%%%%%%%%%%%%%%%%%%%%%%%%%%%%%%%%%%%%%%

\subsection{Auto-coding and auto verification of the software}

In previous sections, we translated traditional observer-based methods into formal properties that should be annotated in the fault detection software.
Annotating the software, software semantics are expressed inside the code, usually in the form of invariant sets to which software variables belong.
In \cite{feron2010control}, it is explained how to annotate a controller software. State variables belong to an invariant ellipsoid.
Each line of the code may or may not change the ellipsoid. If the code makes any change to the ellipsoid, the change is expressed as annotations. The annotations are based on Hoare logic
proposed first by Charlie Hoare~\cite{hoareaxiom69}. For each line of the code, some pre-conditions and some post conditions are provided.
Pre-conditions are supposed to be true before the execution of that line. They might be concluded from previous lines of the code or might be external assumptions. Post-condition must be true after the execution of the line. A verification tool certifies that the post-condition is true if the pre-condition is true if the corresponding line of the code is executed.

There are two main changes that a piece of code may apply to invariant sets.
Firstly, it may perform a linear transformation on the variables. As an example, consider the case in which system states are multiplied by sytem matrix, $A$.
Secondly, the code may concatenate two invariant ellipsoids. This operation leads to a new invariant ellipsoid. As an example, consider the case that bounded inputs are added to bounded system states. In \cite{feron2010control}, S-procedure is used to calculate the resulting ellipsoid.
Annotations express how these two main operations change invariant ellipsoids.

We need to express the annotations in a standard format so that it can be read by standard verification tools.
To unify the annotations, C Specification Language (ACSL)~\cite{baudinacsl08} can be used.
In Table \ref{tab2}, we have a piece of C code that assigns the square of the variable x to
x.

\begin{table}[h]
\centering
\caption{Annotated code sample using ACSL}
\begin{tabular}{| l |} %
\hline
$/*$ \\
  $@ require$ $x\leq 1$\\
  $@ ensures$ $x\leq 2$ \\
$*/$ \\
\{ \\
         x=2*x;\\
\}\\
\hline
\end{tabular}\label{tab2}
\end{table}

Notice the logic predicates $x<=0$ and $x>=0$ right before the piece of code
denoted by the symbols ``@ require'' and ``@ ensures''.
These are annotations expressed in ACSL~\cite{baudinacsl08}.
The ACSL keyword ``require'' denotes a pre-condition.
The ``ensures'' keyword denotes the post-condition.

Another example is given in Table \ref{tab3}. This example corresponds to a simple discrete-time state space system of order one.
The symbols ``@ assumes'' shows an ``external assumption'' on $u$, i.e. the input to the system. Note that the pre-condition `` $x*x<=1$''
might be concluded from previous lines of the code, while the bound on $u$ is not imposed by the code.
If the states are quadratically bounded before the loop, the infinite loop will not change the property and system states must remain
in the original ellipsoid. An example of annotated codes for fault detection and control software is provided in \cite{Tim2013b}.
\begin{table}[h]
\centering
\caption{Annotated code sample using ACSL}
\begin{tabular}{| l |} %
\hline
$/*$ \\
  $@ assumes$ $u*u<1;$\\
  $@ requires$ $x*x<=1$\\
  $@ ensures$ $x*x<=1$\\
$*/$\\
\{\\
         while (1) \{\\
              $ x=0.5*x+0.5*u$;\\
         \}\\
\}\\
\hline
\end{tabular}\label{tab3}
\end{table}

After auto-coding the software, we need to verify the code. For that purpose, Frama-C/WP platform~\cite{framac}~\cite{wp} can read the annotations
produced by ASCL and converts them to logic properties. Such logic statements, then are processed by Why3 tool~\cite{boogie11why3} which converts these properties into a format readable by
the interactive theorem prover PVS~\cite{PVS-CADE92}. PVS is the tool that certifies the correctness of the software.

%%%%%%%%%%%%%%%%%%%%%%%%%%%%%%%%%%%%%%%%%%%%%%%%%%%%%%%%%%%%%%%%%%%%%%%%%%%%%%%%
\section{Conclusions}

In this research we focus on software implementation of the observer-based fault detection methods. Mathematical proofs that support correct operation of the observers are translated in a formal language based on invariant sets and are annotated in software implementation of the observers. Therefore, design specifications and properties can be understood by computer scientists as well as software verification tools. Those tools can certify the original design requirements by tracking software variables that represent observer error states.
Only properties that are supported by mathematical proofs are annotated. Hence, properties that are achieved by trial and error, seem correct without mathematical proof, cannot be verified.
We provide the details of the material that should be documented in the software for three different observer-based fault detection methods. Those methods are proposed for fault detection, isolation and estimation in theory.
The methods are implemented on an experimental lab helicopter system. It is shown that all the methods work fine in practice, detecting a set of  pre-defined fault scenarios. However, software implementation of each method is different. Successful simulation results and even practical tests does not certify that the observer works fine for all possible conditions. However, software verification techniques can certify the properties that are supported by theory and are translated in the formal language.
\bibliographystyle{ieeetr}
\bibliography{biblio}

\appendix
\section{Linear model of the system} \label{model}

According to \cite{Quanser}, the linear model of the system is provided as follows

\begin{eqnarray}
A&=&\begin{pmatrix}
0 & 0 & 0 & 1 & 0 & 0\\
0 & 0 & 0 & 0 & 1 & 0\\
0 & 0 & 0 & 0 & 0 & 1\\
0 & 0 & 0 & 0 & 0 & 0\\
0 & 0 & 0 & 0 & 0 & 0\\
0 & \frac{(2 m_f L_a-m_w L_m) g}{2 m_f L_a^2+2 m_f L_h^2+m_w L_m^2} & 0 & 0 & 0 & 0
\end{pmatrix}, \\
B&=&\begin{pmatrix}
0 & 0\\
0 & 0\\
0 & 0\\
\frac{L_a K_f}{(m_w L_w^2+2 m_f L_a^2)} & \frac{L_a K_f}{m_w Lw^2 +2 m_f L_a^2}\\
 \frac{K_f}{2 m_f L_f} &  \frac{-K_f}{2 m_f L_f}\\
0 & 0
\end{pmatrix}, \\
C&=&\begin{pmatrix}
1 & 0 & 0 & 0 & 0 & 0\\
0 & 1 & 0 & 0 & 0 & 0\\
0 & 0 & 1 & 0 & 0 & 0
\end{pmatrix}.\label{8}
\end{eqnarray}

\begin{figure}[htbp]
\begin{center}
\includegraphics
[scale=0.35,angle=270]{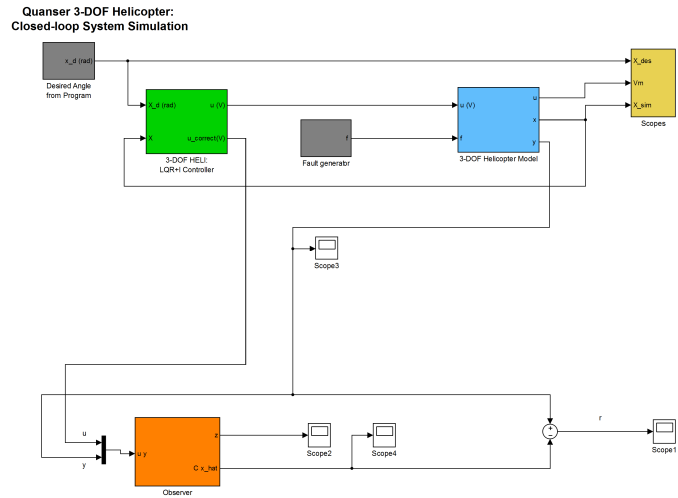}
\end{center}
\caption{Simulink schematic of the closed-loop system together with fault detector}
\label{Pic_simulink}
\end{figure}

\begin{figure*}[htbp]
\begin{center}
\includegraphics
[scale=0.55]{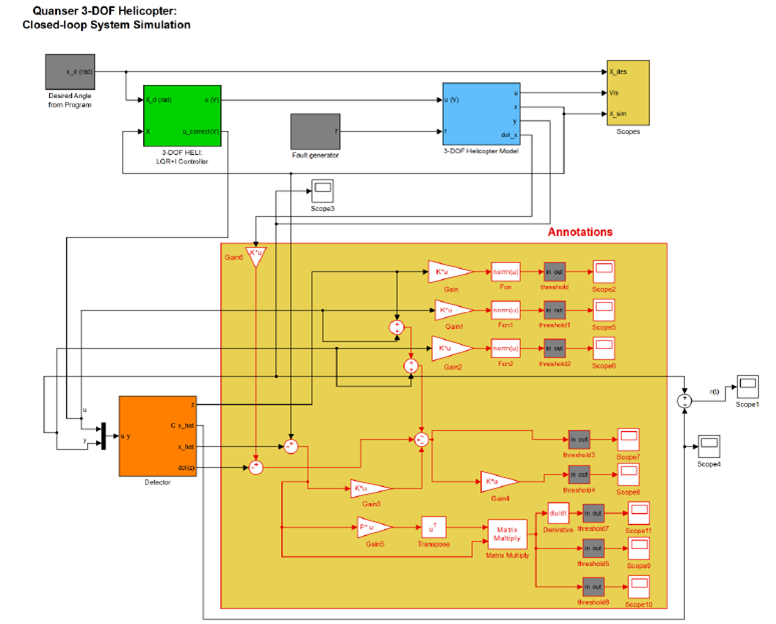}
\end{center}
\caption{Simulink schematic of the closed-loop system together with fault detector and verification annotations}
\label{WithVei}
\end{figure*}

\begin{figure*}[htbp]
\begin{center}
\includegraphics
[scale=0.5]{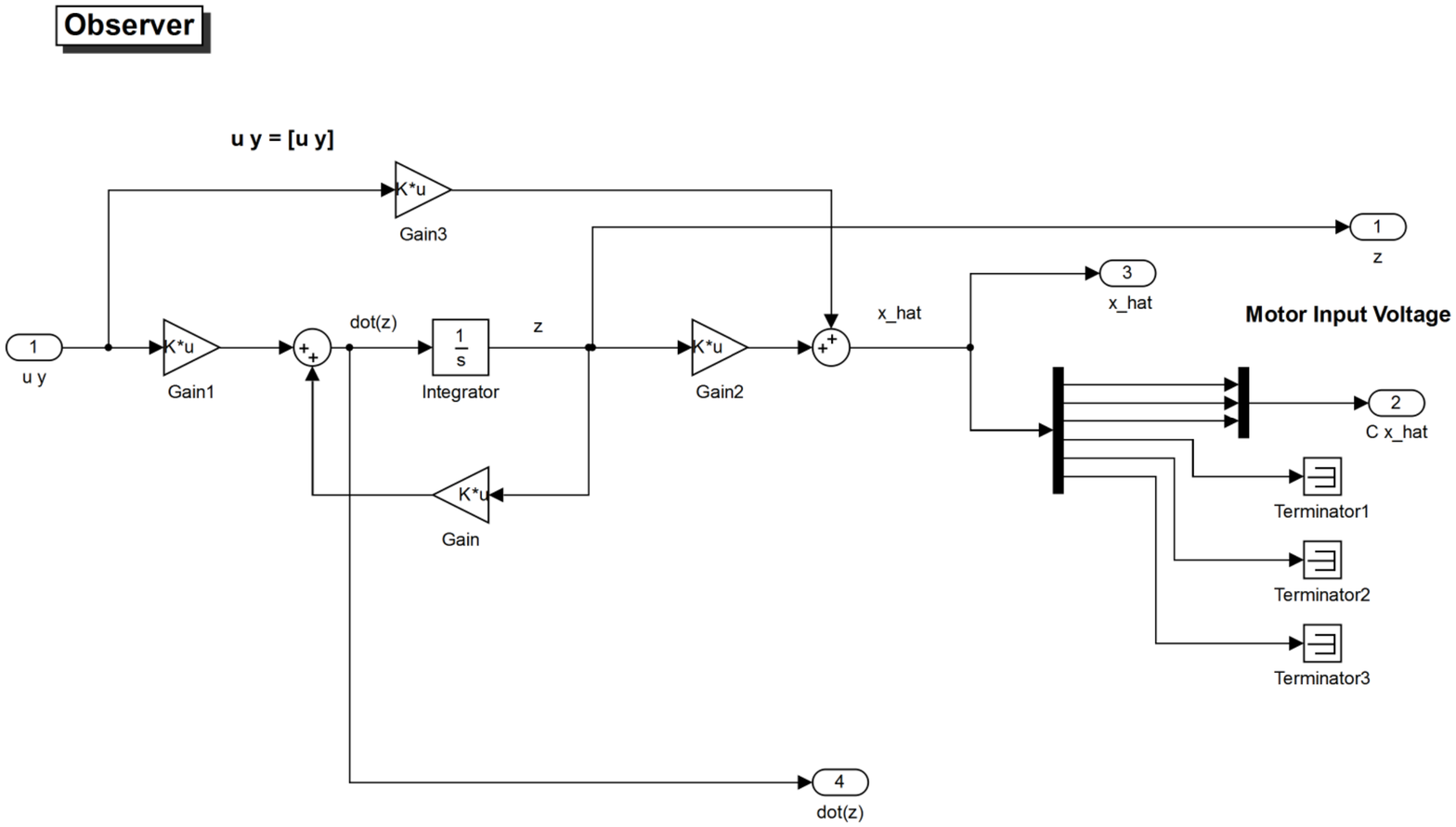}
\end{center}
\caption{Structure of UIO detector}
\label{Observer}
\end{figure*}

\end{document}